\documentclass[breaklinks, colorlinks, onecolumn, useAMS]{aastex63} 
\hypersetup{linkcolor=blue,citecolor=blue,filecolor=cyan,urlcolor=magenta}

\usepackage{enumitem}
\usepackage{mathtools}
\usepackage{graphicx}
   \usepackage{gensymb}
\usepackage{amssymb}
\usepackage{amsmath}
   \usepackage{float}
\usepackage{ulem}
\usepackage{bm}
\usepackage{epstopdf}
\usepackage{color}
\usepackage{subfigure}

\font\fiverm=cmr5

          \font\sixrm=cmr6       

\def\dover#1#2{\hbox{${{\displaystyle#1 \vphantom{(} }\over{
   \displaystyle #2 \vphantom{(} }}$}}

{\catcode`\@=11                                                  
\gdef\SchlangeUnter#1#2{\lower2pt\vbox{\baselineskip 0pt\lineskip0pt    
\ialign{$\m@th#1\hfil##\hfil$\crcr#2\crcr\sim\crcr}}}}           
\def\gtrsim{\mathrel{\mathpalette\SchlangeUnter>}}               
\def\lesssim{\mathrel{\mathpalette\SchlangeUnter<}}

\def\sigt{\sigma_{\hbox{\fiverm T}}}                                
\def\taut{\tau_{\hbox{\fiverm T}}}

\def\teq#1{$\, #1\,$}                         

\def\actionitem#1{\textcolor{blue}{#1}}  
\def\rns{R_{\hbox{\sixrm NS}}}

\def\muB{\mu} 

\def\thetakB{\Theta_{\hbox{\sixrm kB}}}

\def\phik{\phi_{\hbox{\fiverm k}}}

\def\wcyc{\omega_{\hbox{\fiverm B}}}
\def\Bvec{\boldsymbol{B}}
\def\Bvechat{\hat{\boldsymbol{B}}}
\def\Evec{\boldsymbol{E}}
\def\calEvec{\boldsymbol{{\cal E}}}

\def\calEvechat{\hat{\boldsymbol{{\cal E}}}}

\def\kvec{\boldsymbol{k}}
\def\kvechat{\hat{\boldsymbol{k}}}
\def\kvechatS{\hat{\boldsymbol{k}}_{\hbox{\fiverm S}}}
\def\jvechat{\hat{\boldsymbol{j}}}
\def\lvechat{\hat{\boldsymbol{l}}}
\def\rvechat{\hat{\boldsymbol{r}}}
\def\rvechatS{\hat{\boldsymbol{r}}_{\hbox{\fiverm S}}}
\def\rvec{\boldsymbol{r}}
\def\rvecS{\boldsymbol{r}_{\hbox{\fiverm S}}}
\def\rvecShat{\hat{\boldsymbol{r}}_{\hbox{\fiverm S}}}

\def\avec{\boldsymbol{a}}

\def\alphavec{\boldsymbol{\alpha}}

\def\thetaB{\theta_{\hbox{\sixrm B}}}

\def\SigmaB{\Sigma_{\hbox{\sixrm B}}}

\def\kGR{\boldsymbol{k}_{\hbox{\fiverm GR}}}
\def\BGR{\boldsymbol{B}_{\hbox{\fiverm GR}}}
\def\BGRhat{\hat{\boldsymbol{B}}_{\hbox{\fiverm GR}}}

\def\Omegavec{\boldsymbol{\hat{\Omega}}}

\def\muvechat{\hat{\boldsymbol{\mu}}_{\hbox{\fiverm B}}}

\def\taueff{\tau_{{\hbox{\sixrm eff}}}}
\def\Nrec{{\cal N}_{\rm rec}}
\def\PsiS{\Psi_{\hbox{\sixrm S}}}
\def\etaS{\eta_{\hbox{\fiverm S}}}
\def\jSvechat{\hat{\boldsymbol{j}}_{\hbox{\fiverm S}}}
\def\jinftyvechat{\hat{\boldsymbol{j}}_{{\fiverm {\infty}}}}
\def\kSvechat{\hat{\boldsymbol{k}}_{\hbox{\fiverm S}}}
\def\kinftyvechat{\hat{\boldsymbol{k}}_{{\fiverm {\infty}}}}
\def\calESvechat{\hat{\boldsymbol{{\cal E}}}_{\hbox{\fiverm S}}}
\def\calEinftyvechat{\hat{\boldsymbol{{\cal E}}}_{{\fiverm {\infty}}}}
\def\etaorb{\eta}

\newcommand{\vol}[2]{$\,$\rm #1\rm , #2}                 

\shorttitle{Polarization Characteristics of Extended Neutron Star Surfaces}
\shortauthors{Hu et al.}

\begin{document}

\title{INTENSITY AND POLARIZATION CHARACTERISTICS\\ OF EXTENDED NEUTRON STAR SURFACE REGIONS}

\correspondingauthor{K. Hu}
\email{kh38@rice.edu}

\author[0000-0002-9705-7948]{Kun Hu}
\affiliation{Department of Physics and Astronomy - MS 108, Rice University, 6100 Main St., Houston, TX 77251-1892, USA}

\author[0000-0003-4433-1365]{Matthew G. Baring}
\affiliation{Department of Physics and Astronomy - MS 108, Rice University, 6100 Main St., Houston, TX 77251-1892, USA}

\author[0000-0003-0503-0914]{Joseph A. Barchas}
\affiliation{Natural Sciences, Southwest Campus, Houston Community College, 5601 W. Loop S., Houston, Texas 77081, USA}

\author[0000-0002-7991-028X]{George~Younes}
\affiliation{Astrophysics Science Division, NASA Goddard Space Flight Center, Greenbelt, MD 20771, USA}
\affiliation{Universities Space Research Association (USRA) Columbia, Maryland 21046, USA}

\begin{abstract}
The surfaces of neutron stars are sources of strongly polarized soft X rays due to
the presence of strong magnetic fields.  Radiative transfer mediated by electron
scattering and free-free absorption is central to defining local surface anisotropy
and polarization signatures.  Scattering transport is strongly influenced by the
complicated interplay between linear and circular polarizations.  This complexity
has been captured in a sophisticated magnetic Thomson scattering simulation we
recently developed to model the outer layers of fully-ionized atmospheres in such
compact objects, heretofore focusing on case studies of localized surface regions. 
Yet, the interpretation of observed intensity pulse profiles and their efficacy in
constraining key neutron star geometry parameters is critically dependent upon
adding up emission from extended surface regions.  In this paper, intensity,
anisotropy and polarization characteristics from such extended atmospheres, spanning
considerable ranges of magnetic colatitudes, are determined using our transport
simulation.  These constitute a convolution of varied properties of Stokes parameter
information at disparate surface locales with different magnetic field strengths and
directions relative to the local zenith.  Our analysis includes full general
relativistic propagation of light from the surface to an observer at infinity.   The
array of pulse profiles for intensity and polarization presented highlights how
powerful probes of stellar geometry are possible.   Significant phase-resolved 
polarization degrees in the range of 10-60\% are realized when summing over a 
variety of surface field directions.  These results provide an important background
for observations to be acquired by NASA's new IXPE X-ray polarimetry mission.
\end{abstract}


\section{Introduction}
 \label{sec:Intro}

The quasi-thermal soft X-ray emission from isolated pulsars and accreting neutron stars
provides paths to understanding their surfaces and interiors. The surface temperature
distribution can provide clues to the thermal conductivity of the upper crust and help
inform magneto-thermal evolutionary models \citep[e.g.,][]{Vigano13}. The intensity and
polarization signals that are propagated to an observer at infinity are convolutions of
local radiation anisotropies that depend on the global magnetic field structure of the
surface. Such signals likely also depend on whether fully or partially-ionized
atmospheres \citep{Shibanov92,Pavlov94,Zane00,Potekhin04,Suleimanov09} or condensates
\citep{Medin06,Medin07} constitute the outermost layers of the neutron star.

The periodic time traces of soft X-ray intensity provide powerful diagnostics on the
sizes and locales of hot spots, and also on the angles between the magnetic axis and the
observer viewing direction to the spin axis. \cite{Gotthelf10} used such pulse profiles
in a study of PSR J0821-4300 in the supernova remnant Puppis A. Employing a general
relativistic (GR) propagation model, \cite{Perna-2008-ApJ} interpreted the pulse
profiles of the magnetar XTE J1810-197 using a concentric ring polar hot spot that
locally emitted blackbody radiation; they parameterized the radiation anisotropy,
constraining the stellar geometry and mass-to-radius ratio somewhat, yet not profoundly.
\cite{Albano-2010-ApJ} carried out Monte Carlo simulations of spectra and pulse profiles
to constrain the geometric parameters for two transient magnetars XTE J1810-197 and CXOU
J164710.2-455216 in the context of the globally-twisted magnetosphere model.  Their flat
spacetime analysis concluded that XTE J1810-197 is a moderately aligned rotator, with an
angle \teq{\alpha \sim 15^{\circ}-30^{\circ}} between the magnetic and rotation axes,
and a large observer angle \teq{\zeta \sim 150^{\circ}} to the spin axis. In contrast,
they determined that CXOU J164710.2-455216 was an almost orthogonal rotator.
\cite{Bernardini-2011-MNRAS} performed a similar analysis with parameterized radiation
anisotropy for XTE J1810-197 by fully including GR light bending and obtained somewhat
larger magnetic obliquities \teq{\alpha \sim 30^{\circ}-50^{\circ}}, and \teq{\zeta \sim
20^{\circ} - 50^{\circ}}. More recently, \cite{Younes-2020-ApJ} performed flat spacetime
simulations with anisotropic emission to constrain the hot spots' geometry for the
flaring activity from the magnetar 1RXS J1708-40, obtaining \teq{\alpha \sim 60^{\circ}}
and \teq{\zeta \sim 60^{\circ}}.

The use of pulse profiles as probes of stellar parameters for other types of neutron
stars has been extensive over the years. \cite{Riffert-1988-ApJ} explored pulse profiles
for accreting X-ray binaries as a means of discriminating between pencil beam and fan
beam accretion pictures pertaining to columns of plasma inflow and surface impact hot
spots. A similar approach also using the Schwarzschild metric was adopted by
\cite{Muslimov-1998-ApJ} as a path to constraining the \teq{\alpha} and \teq{\zeta}
angles for middle-aged, isolated pulsars.  \cite{Braje-2000-ApJ} studied Doppler
boosting, Shapiro time delays and frame dragging influences of the Kerr metric in
millisecond pulsar (MSP) contexts.  \cite{Psaltis-2014-ApJ} addressed such GR influences
in seeking paths to constrain the mass-to-radius ratio of these rapidly rotating neutron
stars, which have served as core science deliverables for NASA's Neutron Star Interior
Composition Explorer ({\sl NICER}) telescope: see
\cite{Riley-2019-ApJL,Miller-2019-ApJL} for surface ``hot spot'' constraints for MSP
J0030+0451.

The addition of polarization to the suite of neutron star diagnostics enhances the
understanding of the source magnetic geometry.  This observational prospect is on the
near horizon given the very recent launch of the Imaging X-ray Polarimetry Explorer
(IXPE) mission\footnote{IXPE mission details can be found at
\url{https://ixpe.msfc.nasa.gov/}}\citep{Weisskopf16} late in 2021. Polarized X-ray
emission from neutron stars has been explored in numerous previous theoretical studies. 
 For the context of accreting X-ray pulsars, the reader can survey the papers of
\cite{Ventura-1979-ApJ,MB81_ApJ,Meszaros-1988-ApJ} and various articles cited in the
book by \cite{Meszaros-1992} to discern the varied energy-dependent polarization
signatures that define powerful observational probes of accretion geometry. This is
generally the domain of hard X-ray polarimetry, and an introductory demonstration of
diagnostic capability in this band was recently provided by the {\sl X-Calibur}
polarization measurement above 15 keV for the X-ray binary GX 301-2
\citep{Abarr-2020-ApJ}.

Ionized atmosphere models for magnetars constructed by \cite{Ozel-2001-ApJ} and
\cite{HoLai-2001-MNRAS} and a number of later papers generated polarization-dependent
spectra by solving the radiative transfer equations for two photon normal modes,
treating scattering and free-free absorption. Using a sophisticated treatment of the
mode conversion at the vacuum resonance deeper down in the atmosphere,
\cite{Adelsberg06} calculated the polarization signatures from a magnetar hotspot. These
and other studies predicted a predominance of the extraordinary (\teq{\perp}) mode
polarization because it emerges from deeper in the atmosphere, where the plasma is
hotter so that thermal photons are more numerous. Magnetospheric propagation of
polarized light in the birefringent quantum magnetized vacuum is also a key influence
\citep{Fernandez11,Taverna20} in assessing the net polarization signals from magnetars.
As a magnetospheric application, \cite{Taverna17} studied the radiative transfer problem
for magnetar flare emission, identifying strong polarization signatures emerging from
extended closed field line regions of high opacity.

In this paper, results are presented for our scattering-dominated neutron star
atmosphere model, which is based on polarized radiative transfer simulations in the
magnetic Thomson domain. We adopt a complex electric field vector approach that is
numerically expedient when tracking photon polarizations both during atmospheric radiative
transfer and general relativistic propagation through the magnetosphere to an observer at infinity. 
Our results provide more refined determinations of the local surface radiation anisotropy that
significantly improve upon previous invocations, enhancing the modeling of X-ray pulse
profiles from neutron stars. The radiative transfer simulation structure and polarized
injection protocol are summarized in Section~\ref{sec:simulation}.  Intensity and
polarization characteristics from atmospheric slabs at different magnetic colatitudes
are presented in Section~\ref{sec:local_atmos}. Section~\ref{sec:GR_Stokes} details the
implementation of general relativity and the parallel transport of the polarization
signatures. In Section~\ref{sec:extended_atmos}, intensity and polarization signatures
from extended atmospheres are detailed, primarily for monoenergetic photons, emphasizing
entire stellar surfaces, hot polar caps, and also equatorial belts.   It is found, as
expected, that intensity pulse fractions are considerable for polar caps covering
magnetic colatitudes smaller than around \teq{30^{\circ}}, with the fractional
modulation for the entire surface typically being smaller than around 10\%.  Also, as
anticipated, the phase-resolved linear polarization degrees \teq{\Pi_l} of polar caps
can be quite large, up to around 60\%.  Yet, \teq{\Pi_l} drops to the 10-20\% range for
the full surface emission case, somewhat higher than would be realized for random
directions of surface field vectors; this is the imprint of the dependence of Stokes
information on colatitude that the simulation captures. In Section~\ref{sec:context} it
is also shown that these polarization properties prevail when integrating over thermal
X-ray spectra typical of neutron star emission.  The objective of this paper is to
detail the main results and validation of our precision simulation in preparation for
its future application to source observations. An evident strong dependence of the
resultant intensity and polarization pulse profiles on the \teq{\alpha} and \teq{\zeta}
angles indicates the excellent prospects for delivering insightful observational
diagnostics for neutron star geometry parameters using results from our radiative
transfer simulation.

\section{Monte Carlo Simulation of Radiative Transfer}
 \label{sec:simulation} 

This Section summarizes the basic structure of the magnetic Thomson 
scattering simulation {\sl MAGTHOMSCATT}, 
whose design was detailed at length in \cite{Barchas-2021-MNRAS}.  
Our results provide a more sophisticated understanding of neutron
star atmospheres, yielding refinements for predictions of phase-resolved polarization
signals from extended neutron star surfaces that are of great interest for soft
X-ray polarimetry missions like IXPE, and planned future missions 
such as the enhanced X-ray Timing and Polarimetry
mission\footnote{A description of the planned eXTP mission can be found at
\url{https://www.isdc.unige.ch/extp/}.} (eXTP) \citep{Zhang16} and the 
X-ray Polarization Probe (XPP) \citep{Jahoda19}.

\subsection{Summary of Design Elements}
 \label{sec:construction}

The {\sl MAGTHOMSCATT} Monte Carlo simulation treats the transfer of polarized radiation
in magnetized neutron star surface layers. In the code, individual photons are injected
at the base of an atmospheric slab permeated by cold plasma, then they undergo magnetic
Thomson scatterings and eventually escape from the upper planar boundary of the slab.
The distance \teq{s} a photon propagates between scatterings is sampled exponentially
according to Poisson statistics using the total cross section in Eq.~(\ref{eq:sig_mag_Thom_form}), 
and the new propagation direction and polarization is determined by applying the 
accept-reject method to the polarization-dependent differential cross section in 
Eq.~(\ref{eq:dsig_magThom}) below.  The code tracks the complex electric field vectors 
\teq{\calEvec} of individual photons during scatterings, a numerically efficient protocol.
This contrasts previous studies that tracked Stokes parameters \citep[e.g., see][for
white dwarf applications]{Whitney-ApJS91} or normal eigenmodes in solving the radiative
transfer equation for magnetar atmospheres \citep[][]{HoLai-2001-MNRAS,Ozel-2001-ApJ}.
It also differs from the adoption by \cite{Fernandez07,Nobili-2008-MNRAS} of two linear
polarization modes for Monte Carlo simulations of scattering in magnetar magnetospheres.
The complex electric field vector captures the complete polarization information
(linear, circular and elliptical) of a photon, thus it suitably represents photon
polarization near the cyclotron resonance where there is a strong interplay between
circular and linear polarizations due to the induced electron gyration. 
The magnetic scattering formalism in our \teq{\calEvec} vector formalism is 
summarized in Appendix A, and detailed in \cite{Barchas-2021-MNRAS}.
This field vector approach can also be extended to treat dispersive transport in
plasma or the magnetized quantum vacuum, especially near the so-called vacuum resonance
where the contribution of plasma and vacuum dispersion are in strong competition with
each other \citep[e.g., see][]{Meszaros-1992,Lai2003}. 
The photon injection can be adjusted to treat any anisotropy and polarization
configuration at the base of the slab, and Sec.~\ref{sec:injection} details our
injection protocol that captures the precise polarization information from radiation
transfer due to Thomson scattering deep inside the atmosphere. 

\begin{figure*}
\vspace*{0pt}
\centerline{\hskip 0pt 
\includegraphics[height=6.8truecm]{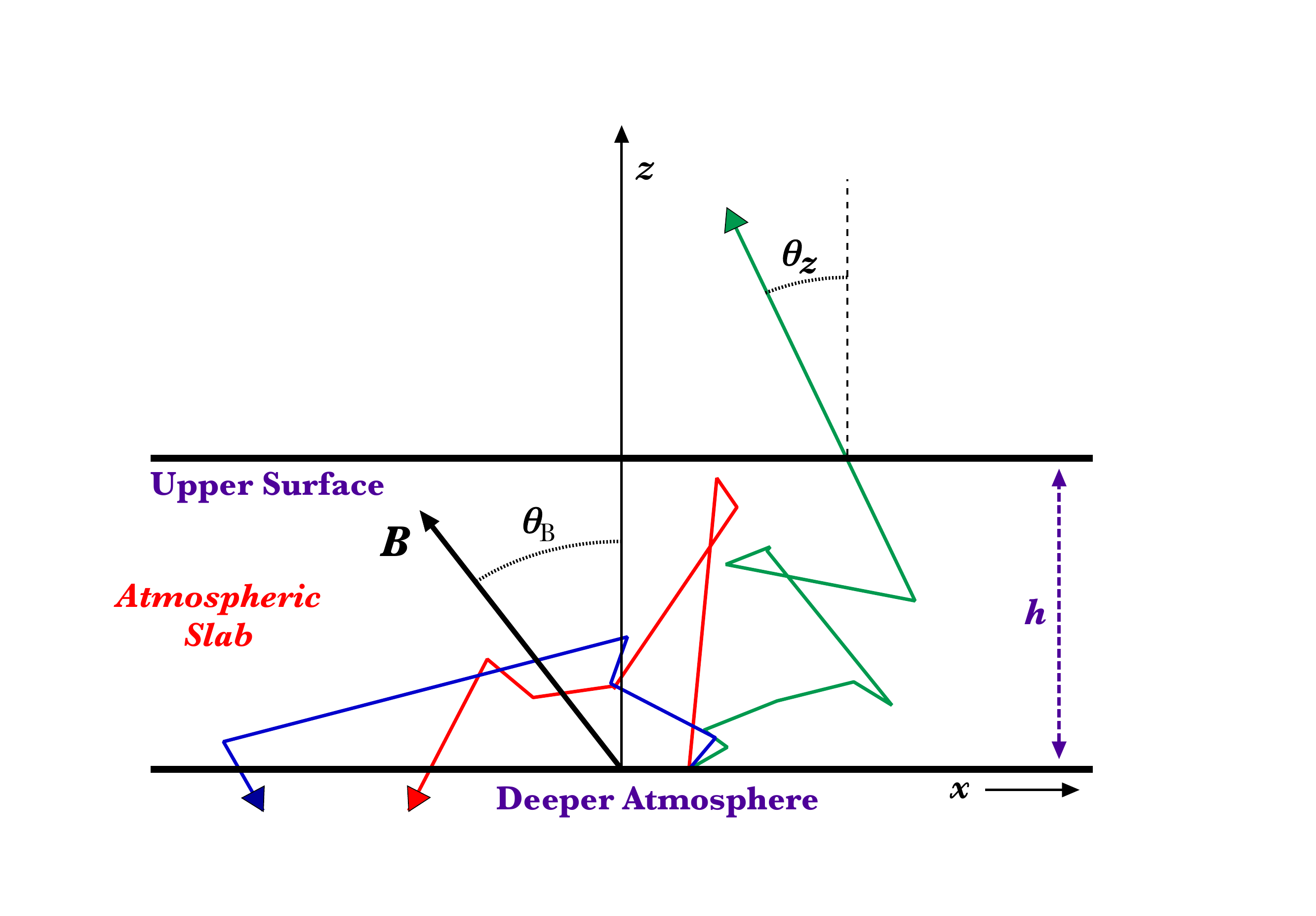}
\hspace{0pt}
\includegraphics[height=7.1truecm]{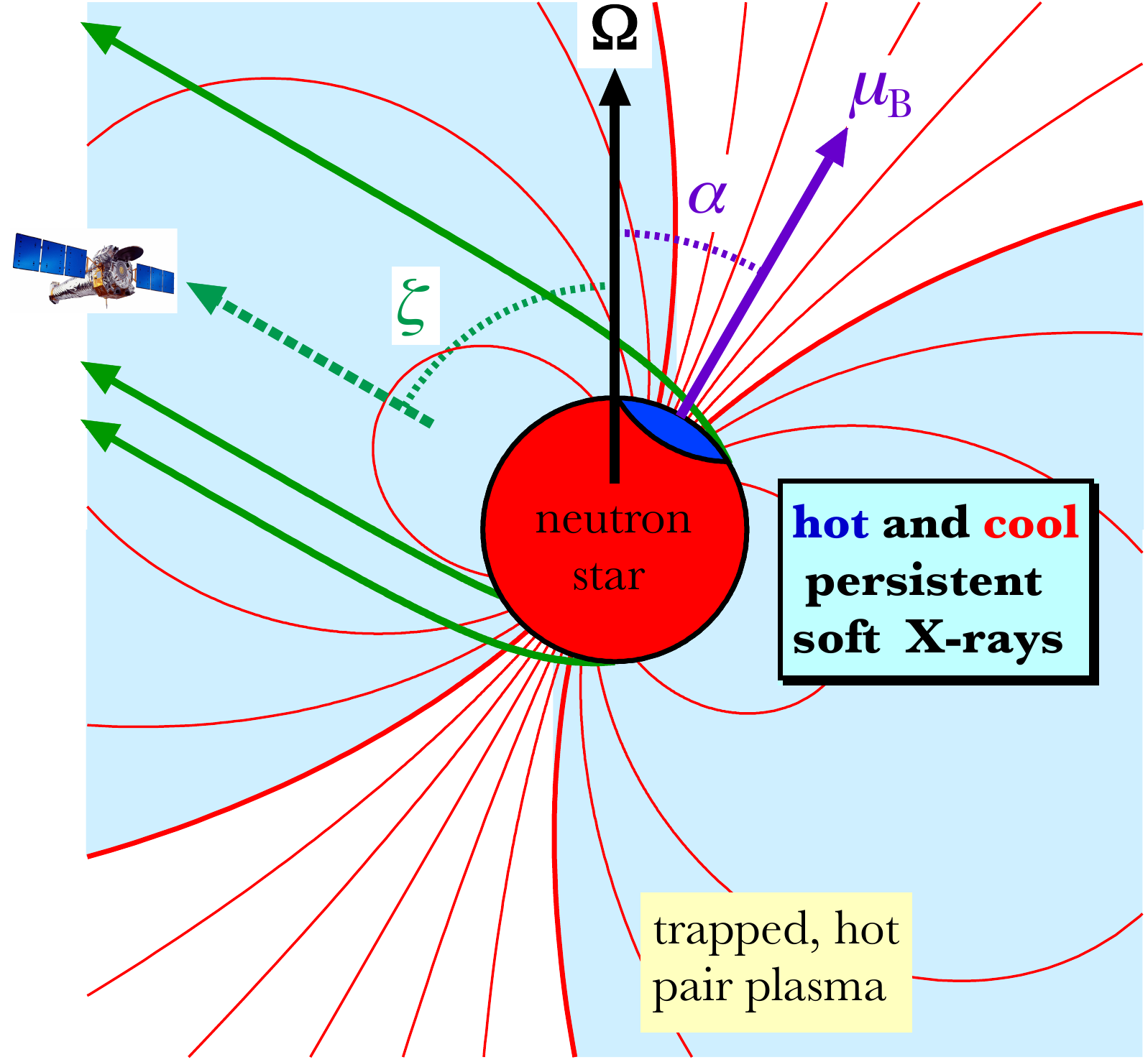}}
\vspace*{-3pt}
\caption{\actionitem{Left:}
Simulation geometry for transfer of photons through an atmospheric slab of height \teq{h} with 
a normal direction along the \teq{z} axis (zenith) and magnetic field, putatively in the \teq{x-z} plane,
at an angle \teq{\thetaB} to the local zenith.  The green trajectory (projected on the 
\teq{x-z} plane) corresponds to a photon that scatters 10 times before exiting the 
top of the slab with zenith angle \teq{\theta_z}, reaching the observer at infinity where its 
Stokes parameter information is recorded.  The blue and red trajectories are for photons 
that scatter and eventually exit the bottom of the slab and are therefore unobservable.
Adapted from \cite{Barchas-2021-MNRAS}.
\actionitem{Right:}
Global geometry of the spinning neutron star, displaying the inclination of the magnetic 
dipole moment \teq{\muvechat} (purple, with dipolar field lines in curved spacetime 
displayed in red) to the rotation axis (signified using the angular velocity vector 
\teq{\Omegavec}) by angle \teq{\alpha}, and a particular viewing perspective (green arrow) 
for a telescope$^{3}$ that is at an angle \teq{\zeta} to \teq{\Omegavec}.
Three coplanar curved trajectories (green) in the Schwarzschild metric are depicted
for photons emanating towards the observer from different magnetic colatitudes on the stellar surface, 
with different local \teq{\theta_z} angles. The blue portion of the surface signifies a putatively hotter 
region and thereby identifies temperature non-uniformity of the stellar atmosphere.
 \label{fig:slab_global_geometry}}
\vspace{-10pt}
\end{figure*}

In the simulation's construction, local slab geometries are embedded in globally extended 
atmospheres and include general relativity effects on the radiation propagating from the 
stellar surface to a distant observer.  Parallel transport of electric field vectors is easily 
tracked when implementing the simulation in curved spacetime; such simplicity is 
not so easily afforded for Stokes parameters.
The geometry of the slab simulation is displayed in Fig.~\ref{fig:slab_global_geometry}
(left), illustrating representative cases of electromagnetic wave transport mediated by
Thomson scatterings.  The slab thickness \teq{h} is captured through optical depths that
serve as proxies: see Eq.~(\ref{eq:tau_eff_def}) below.  Only data for intensity and
polarization of light emergent through the top of the upper atmosphere is recorded when
generating the results of this paper. The deeper portions of the atmosphere where
hydrostatic structure introduces temperature and density gradients
\citep[e.g.,][]{HoLai-2001-MNRAS,Ozel-2001-ApJ} are not modeled here, being deferred to
a future investigation. At the corresponding higher densities, free-free opacity becomes
significant, particularly at photon energies below around 3 keV, thereby modifying the
transport from a purely magnetic Thomson regime. In our study, magnetospheric
propagation of light from different portions of a non-isothermal surface to an observing
telescope\footnote{The Chandra X-ray telescope image is from the CXO resources website
\url{https://www.chandra.harvard.edu/resources/}.}  at infinity is depicted on the right
of Fig.~\ref{fig:slab_global_geometry}.  This schematic diagram also identifies the key
neutron star geometry angles \teq{\alpha} and \teq{\zeta} that can be probed by
phase-resolved pulsation and polarimetric considerations of atmospheric emission from
extended neutron star surface regions: see Sec.~\ref{sec:context}. Although our focus is
on neutron star surface layers, the Monte Carlo technique is quite versatile and can be
used in other neutron star radiative transport environments dominated by Thomson
scatterings, e.g. accretion columns and magnetar magnetospheres.

\subsection{Atmospheric Slab Injection Protocols}
 \label{sec:injection}
 
To render the simulation more accurate and efficient for moderate opacities, the
anisotropy and polarization distributions of the photons injected at the base of the
slab need to represent those appropriate for conditions deeper down in the atmosphere
than {\sl MAGTHOMSCATT} actually simulates.  To achieve this, \cite{Barchas-2021-MNRAS}
presented a high opacity photon simulation configuration, which captured the angular and
the polarization information for photons fully processed by magnetic Thomson
scatterings. This implementation described the asymptotic state of the scattering system
regardless of photons' spatial displacements or boundary conditions.  The angular and
polarization information merely depends on the photon frequency \teq{\omega} and the
angle cosine {\teq{\muB = \cos\thetakB} of the photon wavevector \teq{\kvec} relative to
\teq{\Bvechat}.  This configuration is summarized using empirical approximations for the
Stokes parameters:
\begin{equation}
   I(\muB ) \; =\; \dover{A_{\omega}(\muB )}{\Lambda_{\omega} \,\sigma (\omega, \, \muB )}
   \;\; ,\quad
   A_{\omega}(\muB )\; =\; \dover{3}{2}\, \dover{1 + {\cal A}(\omega )\, \muB^2}{3 + {\cal A}(\omega )} 
   \quad \hbox{for} \quad
   \muB \; =\; \kvechat \cdot \Bvechat \;=\; \cos\thetakB \quad ,
 \label{eq:intensity_anisotropy}
\end{equation}
where \teq{I(\muB)} is the anisotropic intensity of the asymptotic state, and 
\begin{equation}
   \hat{Q}_{\omega} (\muB ) \;\equiv\; 
   \dover{Q}{I}
   \; =\; \dover{{\cal A}(\omega ) \, \left[ \muB^2-1 \right] }{1 + {\cal A}(\omega )\, \muB^2}
   \;\; ,\quad
   \hat{V}_{\omega} (\muB )\;\equiv\; 
   \dover{V}{I} 
   \; =\; \dover{2 {\cal C}(\omega ) \,  \muB }{1 + {\cal A}(\omega )\, \muB^2} 
   \quad , \quad 
   \Pi_{\omega} \; =\; \sqrt{  \bigl(\hat{Q}_{\omega} \bigr)^2 + \bigl( \hat{V}_{\omega}  \bigr)^2 } 
 \label{eq:QV_Pi_omega_mu}
\end{equation}
for the Stokes parameters that are normalized by the intensity \teq{I(\muB)}, and the
derivative {\it total} polarization degree \teq{\Pi_{\omega}}. For local slab
implementations, the coordinate basis used to establish the Stokes parameters can be
chosen to render the Stokes \teq{U} to be zero; for more general geometries this no
longer is possible: see Sec.~\ref{sec:GR_Stokes}.

The re-distribution anisotropy \teq{A_{\omega}(\muB )} constitutes the
frequency-dependent asymptotic probability for re-distributing angles and polarizations
through the scattering's phase matrix \citep{Chou-1986-ApSS}. The intensity
\teq{I(\muB)} satisfies traditional radiation transfer formalism on the sphere
\citep[e.g.,][]{Chandra60}, or equivalently a Boltzmann equation for the photon density
distribution \teq{n_{\gamma} (\mu ) \propto I(\muB )/c}. It is therefore proportional to
\teq{A_{\omega}(\muB )}, yet is scaled by the relative probability for scattering, i.e.
the total cross section \teq{\sigma (\omega, \, \muB )}.  We also introduce an
additional normalization factor \teq{\Lambda_{\omega}} defined by 
\begin{equation}
   \Lambda_{\omega} \; =\; \int_{-1}^1 \dover{A_{\omega}(\mu )}{\sigma (\omega, \, \mu )}\, d\mu 
   \quad \Rightarrow\quad
   \int_{-1}^1 I(\muB )\, d\muB \; =\; 1 \quad ,
 \label{eq:Lambda_def}
\end{equation}
so that the intensity at the point of injection is normalized to unity for a photon.
Note that this differs from the intensity normalization that is adopted for emergent
photons in the graphical illustrations of Section~\ref{sec:local_atmos}, and also those
realized in Section~\ref{sec:extended_atmos}. The interpretations of
\teq{A_{\omega}(\muB )}, \teq{\hat{Q}_{\omega} (\muB )} and \teq{\hat{V}_{\omega} (\muB
)}, together with specific forms for the fitting functions for \teq{{\cal A}(\omega )}
and \teq{{\cal C}(\omega )}, are detailed in Sections 5.2 and 5.3 of
\cite{Barchas-2021-MNRAS}; we note here that both \teq{{\cal A}} and \teq{\cal C} are
purely functions of the ratio \teq{\omega /\wcyc}, where \teq{\wcyc = eB/m_ec} is the
electron cyclotron frequency.

This asymptotic configuration applies to depths sufficiently remote from the upper slab
surface. Thus, \cite{Barchas-2021-MNRAS} proposed that this choice defined an
appropriate protocol for injecting photons at the base of the slab, closely
approximating the true solution at moderate depths \teq{h} for a semi-infinite
atmosphere.   This expeditious approach is of particular importance in the strong field
domain (\teq{\omega/{\wcyc}\ll 1}) where the total cross section for \teq{\perp} mode
photons is significantly reduced relative to the non-magnetic Thomson value \teq{\sigt}.
 In this case, an extremely large \teq{\tau_{\parallel}} (see
Eq.~(\ref{eq:tau_par_per_def}) below) is needed for the intensity and polarization
angular profiles to saturate if one injects isotropic and unpolarized photons at the
base, making for extremely large computation time.  We therefore opt for the anisotropic
and polarized injection in Eqs.~(\ref{eq:intensity_anisotropy})
and~(\ref{eq:QV_Pi_omega_mu})  at the base of an atmospheric patch with arbitrary
\teq{\Bvechat} direction.

To specify the anisotropy, the direction of an injected photon is prescribed by applying
the accept-reject method to the angular distribution of the flux \teq{I(\muB) \cdot
\cos\theta_{z0}}, using a random variate \teq{\xi_{\mu}}, with \teq{0 < \xi_{\mu} < 1}.
Here, \teq{\theta_{z0}} is the angle of the photon wavevector to the slab normal, i.e.,
the zenith direction \teq{\hat{z}}. Thus, \teq{\cos\theta_{z0} = \kvechat \cdot \hat{z}}
and \teq{I(\muB) \cdot \cos\theta_{z0}} defines the photon flux of the intensity
\teq{I(\muB)} passing through a surface element  of unit area within the slab boundary.
An alternative approach, not adopted here, is to select the direction of an injected
photon using a two-step approach, first performing the accept-reject method for the
intensity distribution \teq{I(\muB)}, and then applying the flux weighting to the photon
directions selected by the first step. We explored this possibility and compared these
two selection protocols, determining that they give statistically equivalent results to
a high degree of precision, as expected.

Next, the scaled Stokes parameters \teq{\hat{Q}_{\omega} (\muB )} and \teq{
\hat{V}_{\omega} (\muB )} can be computed using Eq.~(\ref{eq:QV_Pi_omega_mu}), and the
polarization state of the injected photon is determined by comparing the total
polarization degree \teq{\Pi_{\omega}} with a random variable \teq{\xi_{\Pi}} in the
range [0, 1]. If \teq{\xi_{\Pi}\geq\Pi_{\omega}}, the photon is randomly chosen to be in
the \teq{\perp} or \teq{\parallel} mode polarization states, generating a statistically
unpolarized injection. These two orthogonal linear polarization modes in a magnetized
medium are of common usage: the ordinary (O-mode) or \teq{\parallel} polarization state
has an electric field vector \teq{\calEvec} in the plane defined by \teq{\kvechat} and
\teq{\Bvechat}, with the extraordinary state (X-mode, denoted by \teq{\perp}) having
\teq{\calEvec} perpendicular to this plane. In terms of scaled Stokes parameters, both
these polarizations have \teq{\hat{V}=0}, with \teq{\hat{Q}=-1} for the \teq{\perp}
state, and \teq{\hat{Q}=+1} for the \teq{\parallel} state. If instead \teq{\xi_{\Pi}
\leq \Pi_{\omega}}, the electric field components are specified by 
\begin{equation}
  {\cal E}_{\theta} \; =\; \sqrt{ \dover{\Pi_{\omega} + \hat{Q}_{\omega}}{2\Pi_{\omega} \vphantom{\bigl(}} }
  \; ,\quad
  {\cal E}_{\phi} \; =\; \dover{i \hat{V}_{\omega}}{2\Pi \, {\cal E}_{\theta} }
  \; =\; i\, \hbox{sgn}(\hat{V}_{\omega}) \, \sqrt{ \dover{\Pi_{\omega} - \hat{Q}_{\omega}}{2\Pi_{\omega} \vphantom{\bigl(}} } \; .
 \label{eq:Ehat_theta_phi_inj}
\end{equation}
This defines a photon of elliptical polarization, for which the ratio of Stokes \teq{Q}
and \teq{V} equals \teq{\hat{Q}_{\omega}/\hat{V}_{\omega}}. In summary, employing just
two random variates, \teq{\xi_{\mu}} and \teq{\xi_{\Pi}}, the elliptically polarized
photons and the statistically unpolarized \teq{\perp/\parallel} mode photons form a
photon ensemble satisfying the asymptotic photon state described in
Eqs.~({\ref{eq:intensity_anisotropy}) and (\ref{eq:QV_Pi_omega_mu}}).

This anisotropic and polarized injection (AP) protocol establishes a fully scattered
photon ensemble at the base of the atmospheric slab, circumventing the need to simulate
the scatterings deep inside the atmospheres. Only photons emerging from the upper
boundary are recorded, with their number \teq{{\cal N}_{\rm rec}} specified so as to
generate the desired statistics as photons are sequentially injected. We performed tests
using exploratory runs that excluded photons with small scattering numbers, to assess
the role of the effective thickness of the slab.  These tests revealed that such an
exclusion is immaterial for most of the anisotropic and polarized injected simulations
with modest optical depths, except for small \teq{\omega/\wcyc} where the cross section
strongly depends on the polarization and direction of the photon.

\section{Polarization Characteristics from Localized Atmospheres}
 \label{sec:local_atmos}

It is insightful to summarize the character of intensity and Stokes parameter
information emerging from representative locales of atmospheric slabs at different
magnetic colatitudes on neutron stars. This sets the scene for the varied contributions
to the net signals from an extended region of the neutron star surface that propagate to
an observer at infinity.  This section presents an assemblage of such results by
highlighting three particular locales, corresponding to magnetic colatitudes
\teq{\thetaB=0^{\circ}} (pole), \teq{\thetaB=45^{\circ}} (mid latitudes) and
\teq{\thetaB=90^{\circ}} (equator). Furthermore, simulation results are presented for
four select frequencies that bracket the cyclotron frequency, namely \teq{\omega/\wcyc =
0.1, 0.3, 0.99, 3}, spanning high-magnetic, sub-cyclotron domains and quasi-non-magnetic
ones well above \teq{\wcyc}. These frequency and colatitude choices suffice to
illustrate the intricate interplay between linear and circular polarizations.

For expediency, a compromise between precision and efficiency of atmospheric slab
simulations must be achieved.  High opacity runs yield greater precision, and low
opacity runs are faster. Since opacity in magnetic environs is highly dependent on
photon direction, we adopt a versatile measure of the effective optical depth
\teq{\tau_{\rm eff}} in our simulations that captures the opacity and diffusion
information at different frequencies and magnetic colatitudes.  It is defined by
\begin{equation}
   \taueff \; = \; n_e h\, \sigma_{\rm up}(\theta_i \; = \; \thetaB) \; = \;  \dover{\taut}{2} \Bigl\{ \sin^2\thetaB   + \SigmaB (\omega) \left[1+\cos^2\thetaB\right] \Bigr\}\;\; ,\quad
   \hbox{for}\quad \SigmaB (\omega ) \; =\; \dover{\omega^2(\omega^2+\wcyc^2)}{ (\omega^2 - \wcyc^2)^2} \quad.
 \label{eq:tau_eff_def}
\end{equation}
This \teq{\taueff} depth is calculated using the magnetic Thomson cross section 
\teq{\sigma_{\rm up}} for unpolarized (up) radiation (see  \cite{Barchas-2021-MNRAS} 
for details) propagating along the local zenith, and therefore \teq{\taueff} depends 
on the angle \teq{\thetaB} and photon frequency \teq{\omega}.  In particular, two 
special cases, \teq{\thetaB=0^{\circ}} and \teq{90^{\circ} }, can be expressed via
\begin{equation}
   \tau_{\parallel}  \; = \;  n_e h\, \sigma_{\rm up}(\theta_i =0^{\circ})
   \; =\; \taut \,  \SigmaB (\omega ) \quad ,\quad 
   \tau_{\perp}  \; = \;  n_e  h\, \sigma_{\rm up}(\theta_i = 90^{\circ} )
   \; =\; \dover{\taut}{2} \Bigl[1+  \SigmaB (\omega ) \Bigr] \quad .
  \label{eq:tau_par_per_def}
\end{equation}
These represent the optical depths at the magnetic pole and equator along the local
zenith, and were employed in \cite{Barchas-2021-MNRAS}. The effective optical depth
defined in Eq.~(\ref{eq:tau_eff_def}) satisfies \teq{ \taueff \; = \; \tau_{\parallel}
\cos^2\thetaB + \tau_{\perp} \sin^2\thetaB}, thereby representing a tensorial mix of
diffusion properties parallel to and perpendicular to the magnetic field.  For the
purposes of efficiently developing representative results from atmospheric slabs at any
surface locale, we will adopt \teq{\taueff =5} in most of the runs for
Fig.~\ref{fig:slab_ap_iu} and following. Yet we note that higher values of \teq{\taueff}
proved necessary at frequencies below the cyclotron resonance in order to accurately
capture the polarization-dependence of the diffusion and generate accurate measures of
Stokes parameters for radiation emerging from the slabs.

In the local slab simulations, for the purposes of illustration, intensity and
polarization data for the photons escaping above the upper surface are collected in
\teq{n_{\theta} = 90} zenith angle bins and assigned a count number \teq{{\cal N}_j} in
bin \teq{j}. These angular bins are equally spaced on \teq{[0, \pi/2]} with a uniform
width \teq{\Delta\theta_z =  \pi/(2n_{\theta}) = 1^{\circ}}, and they sum over all
photon zenith azimuthal angles \teq{\phi_z}.  Note that this data accounting/binning
choice differs from the protocol for the extended surface analysis of
Section~\ref{sec:extended_atmos}, described therein. The intensity for each
\teq{\theta_z} bin is given by
\begin{equation}
   I_j \; =\; \dover{{\cal N}_j}{\cos\theta_j\sin\theta_j\, \Delta\theta_z\, \Nrec} \quad \Rightarrow\quad
   \sum_{j=1}^{n_{\theta} }  I_j \cos\theta_j\sin\theta_j\, \Delta\theta_z \;=\; 1\quad ,
 \label{eq:Intensity_binning_def}
\end{equation}
where the frequency dependence is not expressed explicitly, since most results in the
paper are for mono-energetic photons.  This can be easily adjusted for thermal spectra
and cases of surface temperature non-uniformity, as desired.  Note the total number
\teq{\Nrec} for recorded photons appears in the denominator, so that the intensity can
be normalized to unity. Thus, intensities are weighted by the flux across the planar
upper surface to the slab.

A suite of results for the three surface locales and four different (monochromatic)
photon frequencies is presented in Fig.~\ref{fig:slab_ap_iu}. For panels with
\teq{\omega/\wcyc > 0.3}, a depth of \teq{\taueff=5} was adopted, collecting data from
\teq{10^8} photons emergent through the top of the slab. For panels with
\teq{\omega/\wcyc=0.3} or \teq{\omega/\wcyc=0.1}, substantially higher \teq{\taueff}
values were chosen so that the simulations employing an anisotropic and polarized
injection give the same results one would acquire with very large depths, namely
\teq{\taueff > 200}.   This ``asymptotic convergence'' was tested to confirm that the
results for polarized injection that are illustrated in Fig.~\ref{fig:slab_ap_iu} are
robust. The intensity \teq{I}, Stokes parameters \teq{Q}, \teq{V} and total polarization
degree \teq{\Pi} are integrated over the azimuthal angles about the local zenith and are
plotted as functions of the zenith angle \teq{\theta_z}. Accordingly, the intrinsic
azimuthal dependence for non-polar slab locales is not explicitly illustrated. The
Stokes \teq{Q}, \teq{V} are normalized to the emergent intensity \teq{I}. The intensity
\teq{I} is normalized using Eq.~(\ref{eq:Intensity_binning_def}), then multiplied by an
extra factor of 0.5 in each panel to improve the visualization.  For comparison, results
from isotropic and unpolarized injection (IU) cases are displayed as traces with various
markers for selected panels. This isotropic and unpolarized injection algorithm, a
computationally less efficient choice,  employs a statistically unpolarized and
isotropic photon ensemble for the injection at the base of the slabs \citep[see Section
3 of][]{Barchas-2021-MNRAS}. We note that at large \teq{\taueff \gg 100} the emergent
Stokes parameter distributions are insensitive to the injection protocol.

\begin{figure}
    \centerline{
    \includegraphics[width=\textwidth]{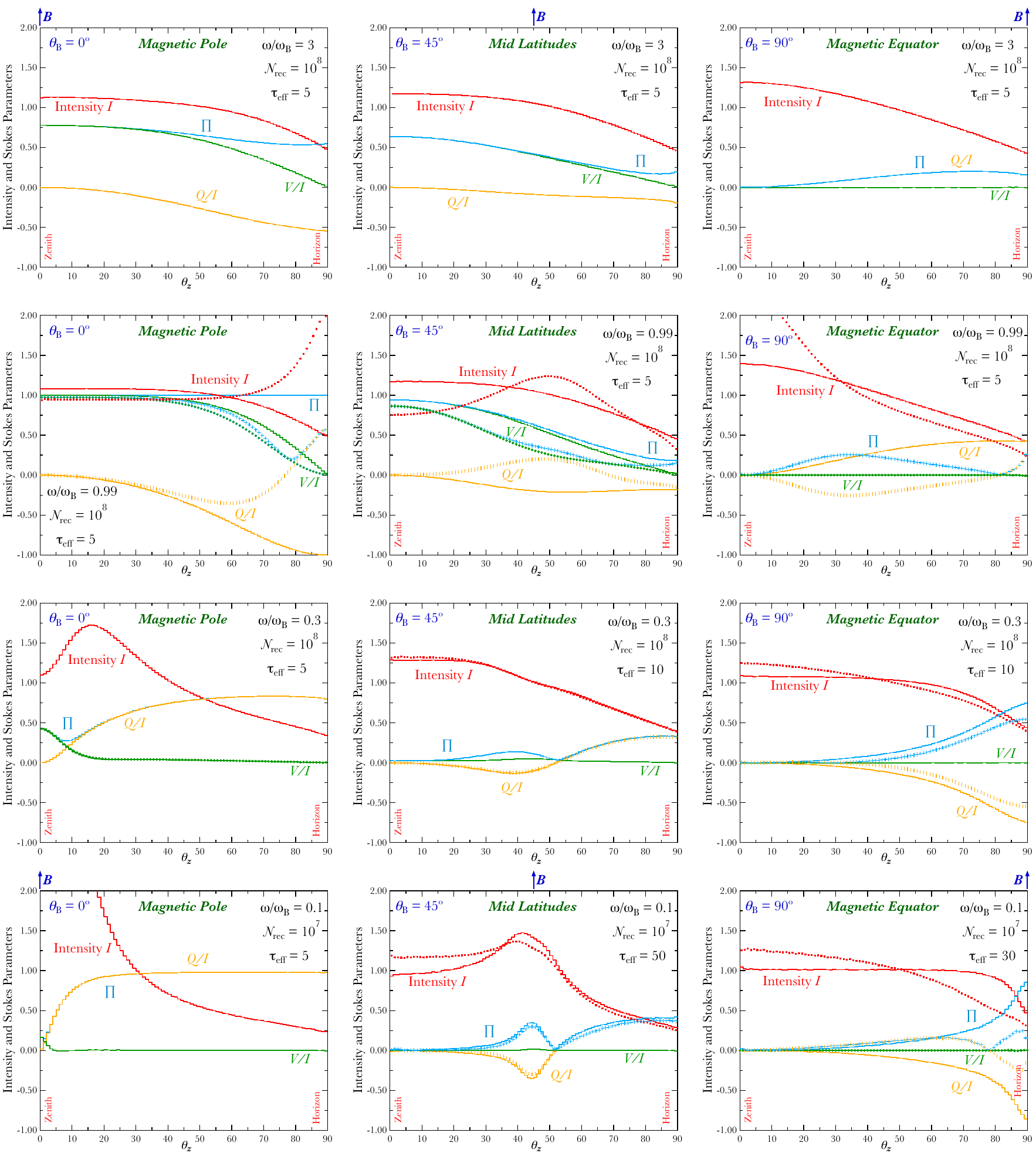}}
    \caption{Angular distributions for intensity \teq{I} (red), Stokes parameters \teq{Q/I} (yellow-orange) 
and \teq{V/I} (green), and total polarization degree \teq{\Pi} (aqua) emergent from the top of the 
atmospheric slabs, as functions of zenith angle \teq{\theta_z} for slab runs at different locales, 
namely three magnetic colatitudes \teq{\thetaB=0^{\circ}, 45^{\circ}} and \teq{ 90^{\circ}}. 
The different rows display results for four different frequency ratios, \teq{\omega/\wcyc = 3, 0.99, 0.3} and \teq{0.1},
from top to bottom.  The optical depth is fixed at \teq{\taueff=5} for \teq{\omega/\wcyc>0.3}. 
Solid histograms display the results for anisotropic and polarized injections,
the preferred AP protocol.  Also, traces with points (squares for \teq{I},
vertical strokes for \teq{Q/I}, dots for \teq{V/I} and crosses for \teq{\Pi}) present results for isotropic 
and unpolarized injection, with all the photons emerging from the upper boundary being recorded 
(not displayed for all panels -- see text for details).}
    \label{fig:slab_ap_iu}
\end{figure}

For \teq{\omega/\wcyc = 3}, the intensity profiles manifest slight anisotropy and
they are insensitive to the emission locales and the injection methods. This
approximates the ``low-field" \teq{\omega\gg \wcyc} domain where the magnetic field
has limited impact on the scattering process and associated diffusion. The
differential cross section then formally approaches the familiar non-magnetic one -- see
the Appendix of \cite{Barchas-2021-MNRAS} for a discussion. At the magnetic pole,
the Stokes \teq{V} is strongest along the zenith direction, indicating significant
circular polarization (+ mode) along \teq{\Bvec}, a direct consequence of the
electron gyration captured in Eq.~(\ref{eq:scatt_accel}). The Stokes \teq{Q} that
describes linear polarization is always negative and strongest along the horizon;
this corresponds to a dominance of the \teq{\perp} mode, always possessing a larger
cross section than the \teq{\parallel} mode (\teq{\hat{Q}=+1}) when \teq{\omega>\wcyc} 
-- see Fig. B1 of \cite{Barchas-2021-MNRAS}. The polarization signatures are considerably 
diminished at non-polar magnetic colatitudes \teq{\thetaB} due to the integration over 
local azimuth: this convolution mixes the particular Stokes parameter information that 
is tied to the field direction and is summarized in Figs.~6 and~7 of
\cite{Barchas-2021-MNRAS}. In particular, the Stokes \teq{V} is always zero at the
magnetic equator, a consequence of exact cancellation of $+$ and $-$ circular
polarization contributions. Observe that the Stokes \teq{Q} parameter switches sign
as the magnetic colatitude \teq{\thetaB} increases in progressing from the pole to
the equator, character that arises at all frequencies. This is a consequence of the
Stokes parameters being calculated in a spherical coordinate system using the local
zenith as a fixed reference direction. Misalignment of the magnetic field to the
local zenith direction changes the Stokes \teq{Q} accordingly, but the dominance of
\teq{\perp} mode polarization locally within the slab doesn't change. The isotropic
injection protocol gives very similar intensity and Stokes profiles at this
frequency, and so they are not displayed in the Figure.

The second row of Fig.~\ref{fig:slab_ap_iu} presents results at
\teq{\omega/\wcyc=0.99}, i.e., very near the cyclotron resonance. In this case,
strong deviations appear between the distributions generated by the polarized and
unpolarized injection protocols; this is expected because of the profound interplay
between linear and circular polarizations in the scattering cross section. These
disparities in emergent \teq{I,Q,V} and \teq{\Pi} profiles between polarized and
unpolarized injection cases are strongest when the viewing perspectives are
orthogonal to the magnetic field direction. This feature is best demonstrated at the
magnetic pole, where the intensity histogram of the isotropic and unpolarized (IU)
injection case realizes a fairly narrow peak at the horizon. The origin of these
disparities is due to the cross section for \teq{\parallel} mode photons not being
resonant when they are propagating orthogonal to \teq{\Bvec}. Therefore
\teq{\parallel}-polarized photons produced by the IU injection protocol can escape
the atmosphere without scattering, thus contaminating the polarization signatures
and forming an intensity peak perpendicular to the magnetic field direction. This
feature disappears for the anisotropic, polarized injection protocol where
scattering is more prolific on average. Test simulations have been performed for 
both injection protocols at \teq{\omega/\wcyc=0.99} with higher optical depth. At
the equator, for example, the convergence of the IU case doesn't occur for
\teq{\taueff<2000}, whereas the convergence of AP injection is clear at
\teq{\taueff=5}, since \teq{\taueff =10,20} runs result in statistically identical
AP distributions. This fact highlights the advantage of the anisotropic and
polarized injection protocol in more efficiently generating the correct Stokes
parameter distributions emergent from the slab.

For the polarized injection case, the intensity profiles for \teq{\omega/\wcyc=0.99}
don't change much from their \teq{\omega/\wcyc=3} counterparts. In contrast, the
polarization information does change with this lowering of frequency, somewhat for
\teq{V} and more so for \teq{Q}. The predominance of the \teq{\perp} mode photons
(\teq{Q < 0}) is now more marked and so the net polarization degree \teq{\Pi} is
enhanced, approaching 100 per cent at the magnetic pole, regardless of the zenith
angle.  This is a key signature of the polarizing influence of driven electron
gyration.  For this polar case, the directional trade-off between circular and
linear polarizations is obvious.    For the equatorial example, Stokes \teq{V} is
again always zero due to the cancellations incurred with azimuthal integration.  As
with the \teq{\omega/\wcyc=3} case, only a modest \teq{\taueff =5} is required at
resonance for the anisotropic, polarized injection to precisely yield the
asymptotic, thick slab signals.

The intensity and Stokes parameter distributions change noticeably when the frequency
drops below the ``equipartition frequency" \teq{\omega/\wcyc = 1/\sqrt{3}\approx 0.577},
where linear polarization cross sections are identical to the Thomson value \teq{\sigt}:
see \cite{Barchas-2021-MNRAS}. This \teq{\omega \ll \wcyc} domain is where the field
strongly impacts the scattering process.  The intensity is beamed close to the
\teq{\Bvec} direction, since the cross section is strongly reduced, i.e.
\teq{\sigma\sim\sigt(\omega/\wcyc)^2}, for photons propagating along the magnetic field
lines, regardless of their polarization. This beaming effect is significant at the
magnetic pole and realizes a peak at \teq{\theta_b\sim\omega/\wcyc}.  The angular
dependence of the peak beaming angle \teq{\theta_b} is associated with the transition
from \teq{\perp} to \teq{\parallel} mode polarization close to the magnetic field,
reflecting the interplay between frequency and angular dependence in the differential
cross sections \citep[see Appendix B of][]{Barchas-2021-MNRAS}. For
\teq{\omega/\wcyc=0.3}, the intensity beaming is muted due to azimuth integration and
flux weighting at emission locales other than the magnetic pole. Yet, for lower
frequencies like \teq{\omega/\wcyc=0.1}, the beaming effect is so strong that it cannot
be completely eliminated by azimuth integration, leaving an intensity enhancement at
\teq{\theta_{z}\sim 45^{\circ}} for the mid-latitude case. The \teq{\parallel} mode
dominates the emergent radiation, while the circular portion is very weak unless viewing
along the \teq{\Bvec} direction. Note that we increase the \teq{\taueff} for the
non-polar cases with \teq{\omega/\wcyc\leq0.3} in order to realize the asymptotic
results expected for very thick slabs. Then, photon diffusion is strong along the
magnetic field at \teq{\omega\ll\wcyc}, while the parameterization of \teq{\taueff} only
captures photon diffusion along the local zenith. Test runs revealed that increasing
\teq{\taueff} above those chosen for the AP runs displayed in all panels in Fig.
\ref{fig:slab_ap_iu} did not incur appreciable changes to any of the \teq{I}, \teq{Q},
\teq{V} and \teq{\Pi} distributions. As with the resonant frequency case, for
\teq{\omega/\wcyc =0.1} the IU run results displayed have not realized such convergence
to the AP distributions, requiring much larger values of \teq{\taueff} to achieve this
asymptotically thick slab state.

\section{Stokes Parameter Transmission in General Relativity}
 \label{sec:GR_Stokes}

This section focuses on the general relativistic (GR) framework for the parallel
transport of the polarization information of light through the magnetosphere in the
curved spacetime of the non-rotating Schwarzschild metric. This choice of metric is
widely applicable to various classes of neutron stars that emit X rays, perhaps with the
exception of millisecond pulsars, and the principal influences can be identified without
imbuing the analysis with the complexities of rotating metrics. Since the magnetic field
structure and ray tracing of light in Schwarzschild spacetime are well-known, the
details of them as implemented in {\sl MAGTHOMSCATT} are summarized in Appendix~B,
therein also detailing our testing protocols for the GR ray tracing. 

\begin{figure*}
\vspace*{0pt}
\centerline{\hskip 0pt 
\includegraphics[height=6.truecm]{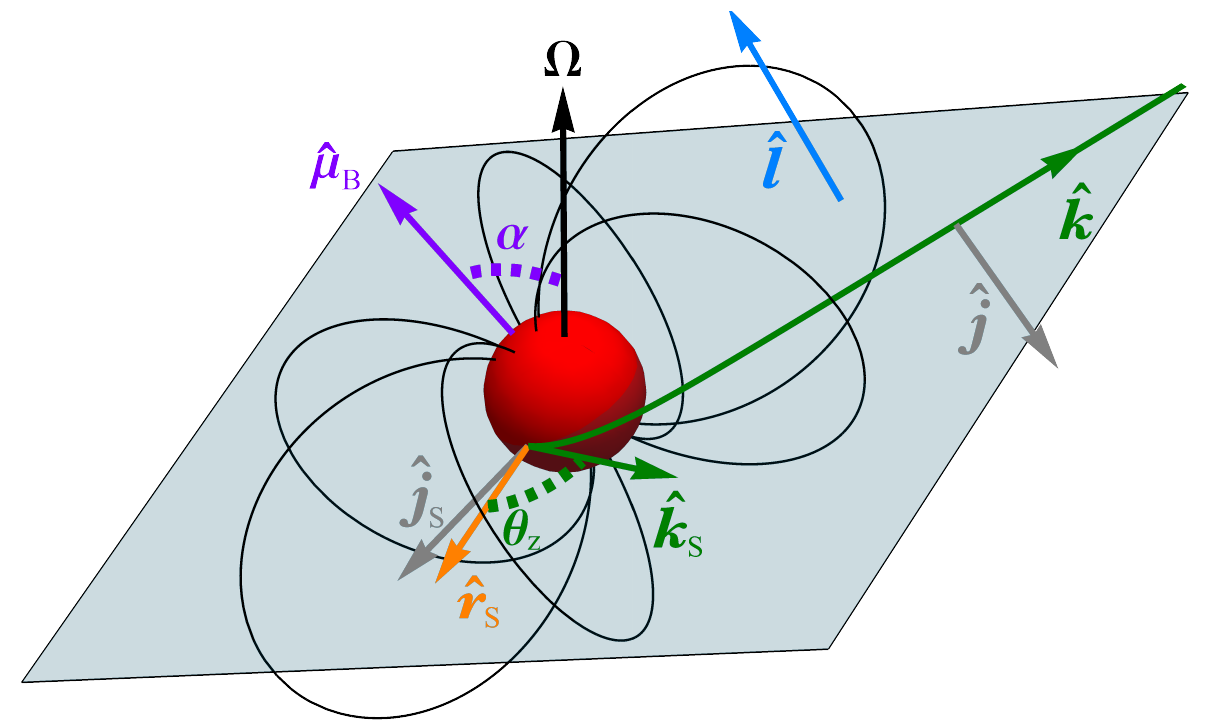}}
\vspace*{-3pt}
\caption{
Global geometry of a spinning neutron star with a inclination angle \teq{\alpha} 
between the angular velocity vector \teq{\Omegavec} and the magnetic dipole 
moment \teq{\muvechat}.  The trajectory plane (light blue) is decided by the cross product 
of the emission locale vector \teq{\rvecShat} and the wavevector \teq{\kSvechat}, 
and it contains the \teq{\jSvechat}, \teq{\kSvechat}, \teq{\jvechat} and \teq{\kvechat} 
vectors (see text).  The angular momentum vector \teq{\lvechat} (blue) of the photon 
trajectory is a constant of motion and is perpendicular to the trajectory plane.
 \label{fig:GR_geometry}}
\end{figure*}

\subsection{Polarization: Parallel Transport -- Mapping Surface to Observer}
 \label{sec:polarize_trans}

In the absence of plasma dispersion and vacuum birefringence, the propagation of
electromagnetic waves outside a neutron star can be well described by the geometric
optics approximation.  The polarization 4-vector \teq{{\cal E}^{\mu}} of the vector
potential undergoes parallel transport along the null geodesic of a light ray and is
always perpendicular to the 4-wavevector \teq{k^{\mu}} \citep[e.g., see Section 22.5 of
][]{MTW-1973}. In a given frame, the time component of \teq{{\cal E}^{\mu}} doesn't
affect the measured electric and magnetic fields, so it is normally set to zero for
simplicity.  The polarization 3-vector (electric field) \teq{\calEvec} lies in the
polarization plane perpendicular to the 3-wavevector \teq{\kvec}. Throughout the
propagation, \teq{\calEvec}  keeps a fixed angle along its trajectory with respect to
the photon orbital angular momentum unit vector \teq{\lvechat} through the sequence of
local inertial frames from the surface to infinity \citep{Pineault-1977-MNRAS}. This
property simplifies the parallel transport of the polarization vector \teq{\calEvec} as
the \teq{\calEvec -\lvechat} plane rotates and remains perpendicular to \teq{\kvec},
with the components of \teq{\calEvec} in the LIF parallel to and perpendicular to the
trajectory plane being constants of the motion. Implementations of this picture can be
found in \cite{Pavlov-2000-ApJ} and \cite{Heyl-2003-MNRAS}.

To specify the polarization at the stellar surface and infinity, we define two groups of
orthogonal basis vectors \teq{\{\jSvechat, \kSvechat, \lvechat\}} and
\teq{\{\jinftyvechat, \kinftyvechat, \lvechat\}} for a photon at the emission (surface)
and observer locales respectively:
\begin{equation}
    \lvechat \; =\; \dover{\rvechat_i\times\kvechat_i}{|\rvechat_i\times\kvechat_{i}|}
    \quad \text{and} \quad
    \jvechat_{i}=\kvechat_{i}\times\lvechat \quad,
 \label{eq:l_j_vechat_def}
\end{equation}
where \teq{i=S} or \teq{\infty}, denoting the stellar surface and the observer at
infinity. The vector \teq{\kvechat_i} is the unit 3-wavevector and \teq{\rvechat_i} is
the unit displacement vector of the photon with respect to the center of the star, with
them generating  \teq{\lvechat}, the constant unit orbital angular momentum vector.  
The \teq{\jvechat_i} vector always lies in the polarization plane.  The differential in
the polarization vector though parallel transport from one LIF to another is mediated by
the affine connection according to \teq{d{\cal E}^{\mu} = - \Gamma^{\mu}_{\nu\lambda}
dx^{\lambda}\, {\cal E}^{\nu}} \citep[e.g., see Sec.~4.9 of][]{Weinberg72}.  The
cumulative increment of this differential along the trajectory is described simply by a
rotation through a single angle, \teq{\etaS - \theta_z}, representing the bending angle
of the photon momentum vector from surface to infinity; this is summarized in
\cite{Pineault-1977-MNRAS}. Here, \teq{\etaS} defines the angle around the stellar
surface in its LIF from the photon emission point \teq{\rvecS} to the point \teq{\rns
\rvechat_{\infty}} on the surface radially in the direction to the observer. Since the
photon emerges from the atmosphere in the direction \teq{\kSvechat = \kGR (\Psi =\PsiS)
/\vert \kGR \vert} at an angle \teq{\theta_z} to its local zenith, its  \teq{ \jSvechat
- \lvechat } plane of polarization lies at an angle \teq{\etaS - \theta_z} to the plane
normal to \teq{\rns \rvechat_{\infty}} in the LIF at the surface. Simple geometry on the
sphere relates the various unit vectors:
\begin{equation}
     \jinftyvechat \; =\; \sin{(\etaS-\theta_z)}\kSvechat+\cos(\etaS-\theta_z )\jSvechat 
     \quad\text{and}\quad
     \kinftyvechat \; =\; \cos{(\etaS-\theta_z)}\kSvechat-\sin(\etaS-\theta_z )\jSvechat
     \quad .
 \label{eq:k_j_infty_hat}
\end{equation}
It is convenient to resolve the polarization vector \teq{\calESvechat} in terms of 
components parallel to \teq{\lvechat} and \teq{\jSvechat} at the stellar surface:  
\teq{\calESvechat = {\cal{E}}_l \lvechat + {\cal{E}}_j \jSvechat}.  The components of 
\teq{\calESvechat} projected onto the  \teq{\jinftyvechat - \lvechat} plane are then 
\teq{{\cal{E}}_l \lvechat} and \teq{{\cal{E}}_j \jinftyvechat}.  One quickly concludes 
that the polarization vector measured by an observer at infinity is then simply
\begin{equation}
    \calEinftyvechat \; =\; {\cal{E}}_l \lvechat+{\cal{E}}_j \jinftyvechat \;\; ,
    \quad \hbox{with}\quad
    {\cal{E}}_l \; =\; \calESvechat\cdot\lvechat
    \quad,\quad
    {\cal{E}}_j\; =\; \calESvechat\cdot \jSvechat \quad ,
  \label{eq:calE_l_J}
\end{equation}
with Eq.~(\ref{eq:k_j_infty_hat}) capturing the rotation information.
The protocols we employed for testing the polarization transport implementation 
are discussed in Appendix C.

Polarization is conveniently measured at infinity, for any rotational phase,
using Cartesian coordinate basis vectors defined with respect to the plane 
defined by \teq{\kinftyvechat} and the stellar angular momentum vector 
\teq{\Omegavec} of the star, i.e.,
\begin{equation}
    \hat{y }\;=\; \dover{ \Omegavec\times\kvechat_{\infty} }{\vert  \Omegavec\times\kvechat_{\infty} \vert}
    \quad\text{and}\quad
    \hat{x}\;=\; \hat{y} \times \kvechat_{\infty} \quad ,
    \label{eq:x_y_obs_polarize}
\end{equation}
so that obviously \teq{\kvechat_{\infty}=\hat{z}}.  For \teq{\cos\zeta = \kvechat_{\infty} \cdot \Omegavec}, 
one then deduces that \teq{\Omegavec = - \sin\zeta\, \hat{x} + \cos\zeta\, \hat{z}}. 
The corresponding components of the electric field vector are then 
\teq{{\cal{E}}_{x} = \calEinftyvechat\cdot \hat{x}} and
\teq{{\cal{E}}_{y} = \calEinftyvechat\cdot \hat{y}}, and these can be inserted in 
Eq.~(\ref{eq:Stokes_polar_def}) to compute the Stokes parameters. 
The photon electric field component parallel to the \teq{\hat{x}} contributes to a positive \teq{Q} value 
while the component parallel to the \teq{\hat{y}} contributes to a negative \teq{Q}.

\subsection{Observables for rotating neutron stars}
 \label{sec:rotation_modulation}

In detailing how the photon momentum and polarization vectors are transported from the 
stellar surface to infinity,  the spin axis \teq{\Omegavec} and the observer's direction 
\teq{\kinftyvechat} are fixed.  Yet this propagation is independent of the magnetic moment 
orientation \teq{\muvechat} which rotates as the star spins.  The zero of phase 
\teq{\Omega t=0} is chosen to be when the magnetic axis \teq{\muvechat} is coplanar 
with \teq{\Omegavec} and \teq{\kinftyvechat}.  The vector \teq{\kinftyvechat}, characterized 
by a viewing angle \teq{\zeta} such that \teq{\cos\zeta = \Omegavec \cdot \kinftyvechat}, 
corresponds to an infinite number of trajectories directed to a specific observer.
For a given \teq{\etaS}, these trajectories form a cylinder at infinity that is constricted 
down near the surface where the intersection of this trajectory surface with the star is a circle. 
This circle samples a variety of magnetic colatitudes and longitudes, values that are modulated 
as the star rotates.  The instantaneous viewing angle \teq{\theta_v} between the line of sight, 
\teq{\kinftyvechat}, and the magnetic axis, \teq{\muvechat}, can be expressed in relation to 
\teq{\Omegavec} via a spherical triangle relation \citep[see section 4.3 of][]{Hu-2019-MNRAS}:
\begin{equation}
    \cos\theta_v \;\equiv\; \kinftyvechat\cdot \muvechat 
       \;=\; \cos\alpha \cos\zeta + \sin\alpha \sin\zeta \cos \Phi \quad ,
  \label{eq:cos_theta_v}
\end{equation}
where \teq{\alpha} is the inclination angle, \teq{\zeta} is the observer angle and
\teq{\Phi=\Omega t} is the rotation phase. This relation controls how the atmospheric
emission information at a given surface locale maps to observers at all rotational
phases. When a photon escapes the atmosphere at a certain locale, its local
\teq{\calESvechat}, \teq{\kSvechat}, \teq{\lvechat} and \teq{\jSvechat} vectors are
specified in Cartesian form. In order to treat a star with arbitrary inclination angle
and rotational phase, we rotate the star and all these vectors first through
\teq{\alpha} from the local magnetic coordinate description identified in Appendix B,
and then through the phase \teq{\Phi=\Omega t}. The rotated vectors at the surface are
then transported to infinity using Eqs.~(\ref{eq:k_j_infty_hat}) and~(\ref{eq:calE_l_J}).

In the simulation, we collect all the photons that propagate to infinity after they
emerge from the neutron star atmospheric locale where their radiative transport is
computed.  This is equivalent to having a large number of observers spanning the entire
sky.   Information captured by observers with the same \teq{\zeta} but different azimuth
angles \teq{\Phi} about the spin axis \teq{\Omegavec} essentially corresponds to the
information collected by the same observer of different rotational phases \teq{\Phi \to
-\Omega t}. In this way, we circumvent the need to simulate different rotational
configurations and are able to collect information of all the rotational phases from a
single simulation run, subsequently assigning it to a rotational phase for each binned
value of \teq{\zeta}. This is the protocol we adopt for the sky maps illustrated in
Sec.~\ref{sec:extended_atmos}.

\section{Polarization Characteristics from Extended Atmospheres}
 \label{sec:extended_atmos}

In this Section, the ensemble intensity and polarization signals are computed 
for extended surface regions.  For each injected photon, the emitting locale is uniformly 
sampled from the prescribed surface region, which for simplicity is chosen 
to be a spherical rectangular patch bounded by fixed magnetic colatitudes \teq{\theta}
and longitudes \teq{\phi} defined by the GR dipole construction detailed in Appendix B.
Throughout the paper, \teq{M = 1.44M_{\odot}} and \teq{\rns = 10^6} cm are adopted. 
Emission locales inside the spherical rectangle patch are sampled using 
\begin{equation}
   \cos \theta  \; =\;  \cos{\theta_l}+(\cos{\theta_u} - \cos{\theta_l})*\xi_{\theta}
   \quad \text{and} \quad
   \phi \; =\;  \phi_l+ (\phi_u-\phi_l)*\xi_{\phi} \quad ,
 \label{eq:patch_sampling} 
\end{equation}
where \teq{\theta_l, \theta_u, \phi_l}, and \teq{\phi_u} are the boundary coordinates of
the patch, and \teq{\xi_{\theta}} and \teq{\xi_{\phi}} are random variates. This
sampling is therefore uniform in solid angle within the spherical rectangle, and is
routinely adaptable to treat temperature gradients that are inferred for most neutron
star systems.  Throughout, azimuthal symmetry is presumed, corresponding to
\teq{\phi_l=0} and \teq{\phi_u=2\pi}. For each photon, the local atmospheric transport
simulation was performed as described in Sec.~\ref{sec:local_atmos} and then propagated
to infinity as detailed in Sec.~\ref{sec:GR_Stokes}, recording its direction in the sky
and its Stokes parameters.

\subsection{Entire Surface}
 \label{sec:entire_surface}

The intensity \teq{I}, Stokes parameters \teq{Q/I, U/I} and \teq{V/I} from a uniformly
emitting star, where 5\teq{\times 10^9} photons are recorded for atmospheric slab
simulations distributed randomly across the whole surface, are illustrated as sky maps
in Figure.~\ref{fig:skymap_w0-1}. The coordinates in these maps are the rotational phase
\teq{\Phi = \Omega t} on the x-axis and the observer's viewing angle \teq{\zeta}
relative to the spin axis on the y-axis. Accordingly, the maps are akin to those used in
models of gamma-ray pulsars \citep[][]{Harding-2015-ApJ,Harding-2017-ApJ}.  The mass of
the star is fixed at \teq{1.44M_{\odot}} which corresponds to a surface radial parameter
\teq{\PsiS\approx0.425}. For each panel, the photon frequency \teq{\omega} is fixed so
that \teq{\omega/\wcyc=0.1} in the LIF at the magnetic poles, with a corresponding value
of \teq{\omega/\wcyc=0.175} at the magnetic equator. From left to right, each column in
Figure.~\ref{fig:skymap_w0-1} displays results for inclination angles \teq{\alpha
=15^{\circ}, 45^{\circ}} and \teq{75^{\circ}}. The sky maps are binned with both
\teq{\zeta} and \teq{\Phi} resolutions of \teq{3^{\circ} }, hence the average photon
count per pixel is around \teq{3\times10^4}. 

\begin{figure}
    \centerline{
    \includegraphics[width=\textwidth]{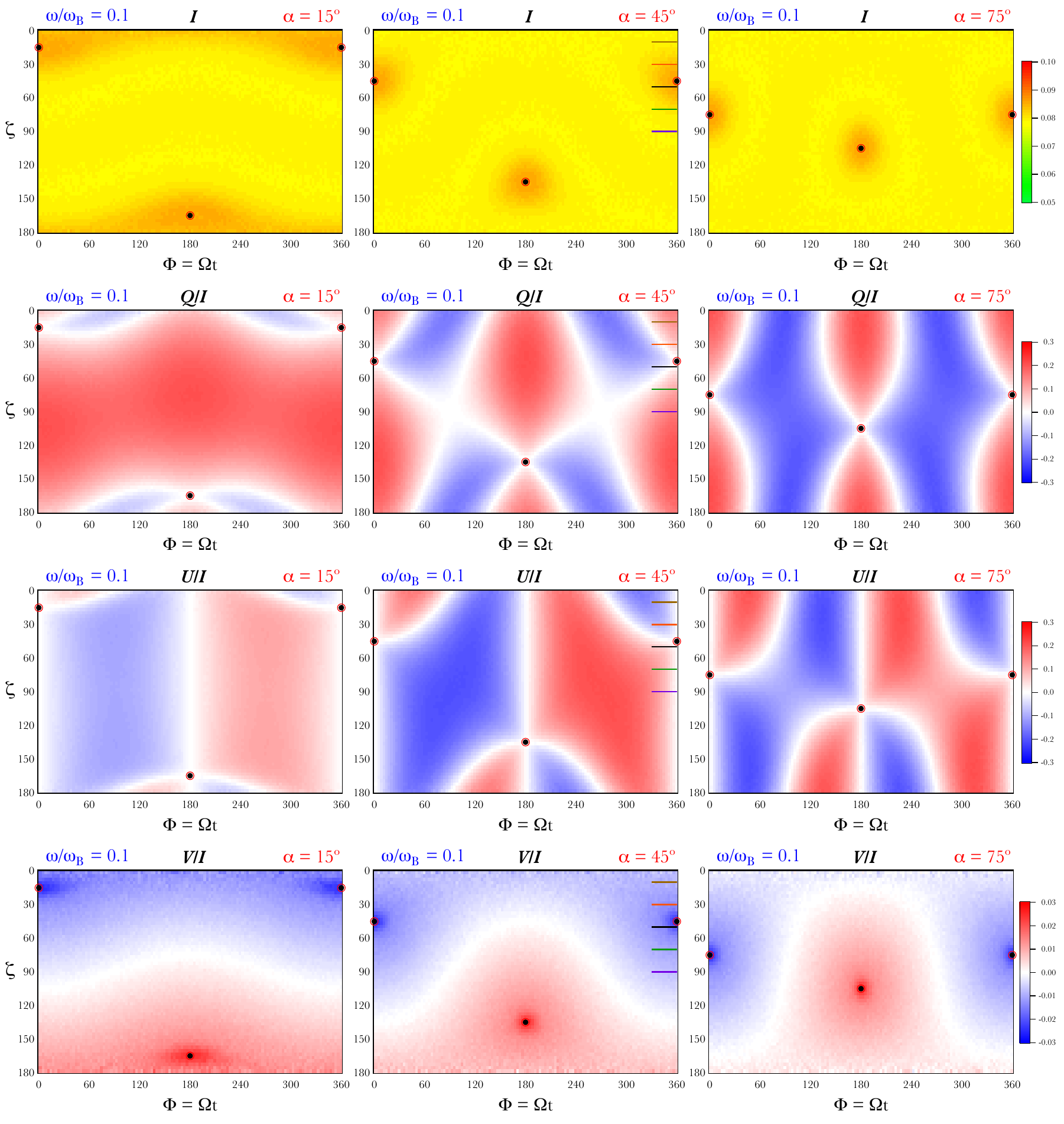}}
    \caption{
    Intensity and polarization sky map distributions obtained via general relativistic propagation 
    of simulated atmospheric signals from a uniformly emitting star.  The sky distributions are 
    plotted as functions of the viewing angle \teq{\zeta} and rotation phase \teq{\Phi = \Omega t}, 
    with scales indicated in the color legend bars at the right of each row.
    The inclination angle of the star \teq{\alpha} is set to \teq{15^\circ, 45^\circ} and \teq{75^\circ} 
    for the columns from left to right, and the mass is \teq{1.44M_{\odot}}(\teq{\PsiS\approx0.425}).
    The frequency of the emitted photons is fixed with a ratio \teq{\omega/\wcyc = 0.1} in the LIF
    at the magnetic poles, and is higher at all other colatitudes where the surface field is lower.
    The number of the recorded photons is \teq{{\cal N}_{\rm rec} = 5 \times 10^9} for each panel.
    The black dots in each panel mark the \teq{(\zeta ,\, \Phi )} combinations 
    for which the observer is instantaneously viewing along the magnetic axis  \teq{\muvechat}. 
    In the \teq{\alpha = 45^{\circ}} panels, horizontal lines mark the \teq{\zeta} values used 
    to assemble the pulse profiles displayed in Fig.~\ref{fig:pulse_profile_0-1_wholestar}
    with the same color coding.} 
 \label{fig:skymap_w0-1}
\end{figure}

The intensity maps display moderate anisotropy with an average value equalling
\teq{1/(4\pi)\sim 0.08}, the intensity normalization. The intensity is maximized when
the observer's line of sight is parallel or antiparallel to the magnetic axis
\teq{\muvechat}. These viewing directions are marked by circled dots in all the sky
maps. The maximum intensity is only around 6\% greater than the average value. The
maxima are due to both the local beaming of radiation emerging from the surface and the
GR lensing of light. The local beaming of the emitted photons is controlled by the local
field direction and strength in the atmosphere: see Fig. \ref{fig:slab_ap_iu} for
examples. Yet, since a specific observer detects photons from regions encapsulating a
range of field strengths and widely differing directions, the intensity variations
evident in from the array in Fig.~\ref{fig:slab_ap_iu} are moderated considerably. In
addition, the GR light bending effect increases the visible surface of the star
\citep[e.g. see][]{Nollert-1989-AA} by 74\% for our choice of \teq{M} and \teq{\rns},
providing access to the far side of the star, so that the intensity variations are
further changed. When \teq{M\to 0} (i.e., \teq{\PsiS=0}) and the GR influences are
removed, the intensity variation is enhanced to around 30\%, with maxima for viewing
directions roughly perpendicular to \teq{\muvechat}, i.e., over equatorial regions of
larger surface area.

The \teq{Q/I} and \teq{U/I} maps are displayed in the second and the third row of
Fig.~\ref{fig:skymap_w0-1} respectively. Here the Stokes parameters are measured using
axes defined in Eq.~(\ref{eq:x_y_obs_polarize}), i.e., with respect to the plane defined
by the observer direction \teq{\kinftyvechat} and the stellar rotational vector
\teq{\Omegavec}. This coordinate choice differs from that adopted for the panels in
Fig.~\ref{fig:slab_ap_iu} where Stokes parameters are measured with respect to the plane
defined by the photon emission direction and the normal to the atmospheric slab
(zenith). The Stokes \teq{Q/I} is mainly positive for stars with small \teq{\alpha}.
This is because the \teq{\parallel} mode polarization dominates the outgoing emission at
the selected frequency range, and the projection of the magnetic axis on the observer's
sky is almost parallel to the reference axis \teq{\hat{x}} defined in
Eq.~(\ref{eq:x_y_obs_polarize}) for small \teq{\alpha}. As the inclination angle
\teq{\alpha} increases, the \teq{Q/I} sky map pattern is noticeably more complicated,
and the \teq{Q/I} values become sensitive to \teq{\zeta} and \teq{\Phi}. This modulation
arises because the projection of the magnetic axis on the sky oscillates around the
\teq{\hat{x}} axis as the star rotates. The \teq{U/I} values are generally non-zero
unless the projection of the magnetic axis on the observer's sky is parallel or
antiparallel to one of the reference axes (\teq{\hat{x}, \hat{y}}) for the Stokes
parameters. The typical linear polarization degree is around 0.2, which is slightly
higher than the flat spacetime values in Fig. 5.5 of \cite{Barchas17}. The \teq{V/I}
results are shown in the last row. The typical \teq{V/I} value is around 0.02, which is
much smaller than the typical linear polarization degree. The right-handed and
left-handed photons dominate the circular polarization when the observer is
instantaneously looking down the south or the north magnetic pole. The value and the
sign of the \teq{V/I} reflect the convolution of a wide variety of magnetic field
directions on the stellar surface, and is naturally expected to be small. Due to the
hemispherical and the azimuthal symmetries of the dipole configuration that is uniformly
sampled by photons, the Stokes parameters in all the maps satisfy the following relations:
\begin{eqnarray}
   I(\zeta, \Phi) \; =\; I(180^{\circ}-\zeta, 180^{\circ}-\Phi) 
   & \; =\;  & I(180^{\circ}-\zeta, 180^{\circ}+\Phi) \; =\; I(\zeta, 360^{\circ}-\Phi),\nonumber \\ 
   \hat{Q}(\zeta, \Phi) \; = \; \hat{Q}(180^{\circ}-\zeta, 180^{\circ}-\Phi) 
   & = & \hat{Q}(180^{\circ}-\zeta, 180^{\circ}+\Phi) \;=\; \hat{Q}(\zeta, 360^{\circ}-\Phi), \nonumber \\ [-7pt]
 \label{eq:IQUV_sym}\\[-7pt]
   \hat{U}(\zeta, \Phi) \; =\; \hat{U}(180^{\circ}-\zeta, 180^{\circ}-\Phi) 
   & = &-\hat{U}(180^{\circ}-\zeta, 180^{\circ}+\Phi) \;=\; -\hat{U}(\zeta, 360^{\circ}-\Phi), \nonumber\\ 
   \hat{V}(\zeta, \Phi) \; = \; -\hat{V}(180^{\circ}-\zeta, 180^{\circ}-\Phi) 
   & = & -\hat{V}(180^{\circ}-\zeta, 180^{\circ}+\Phi)\; =\; \hat{V}(\zeta, 360^{\circ}-\Phi). \nonumber
\end{eqnarray}
Note that a minus sign appears in the \teq{V/I} relation, being 
related to the direction of the magnetic field.  The linear polarization present here, 
with typical degrees of \teq{\Pi_l \sim 20}\%, should be detectable by IXPE \citep{Weisskopf16}
in sources providing sufficient photon count statistics, for example select magnetars.
The vacuum birefringence effects of polarization transport in the magnetized quantum vacuum 
will largely preserve the polarization of photons when they propagate through the 
magnetosphere \citep[][]{Heyl-2003-MNRAS}; their treatment is deferred to future investigations.

\begin{figure}
 \centerline{
   \includegraphics[width=\textwidth]{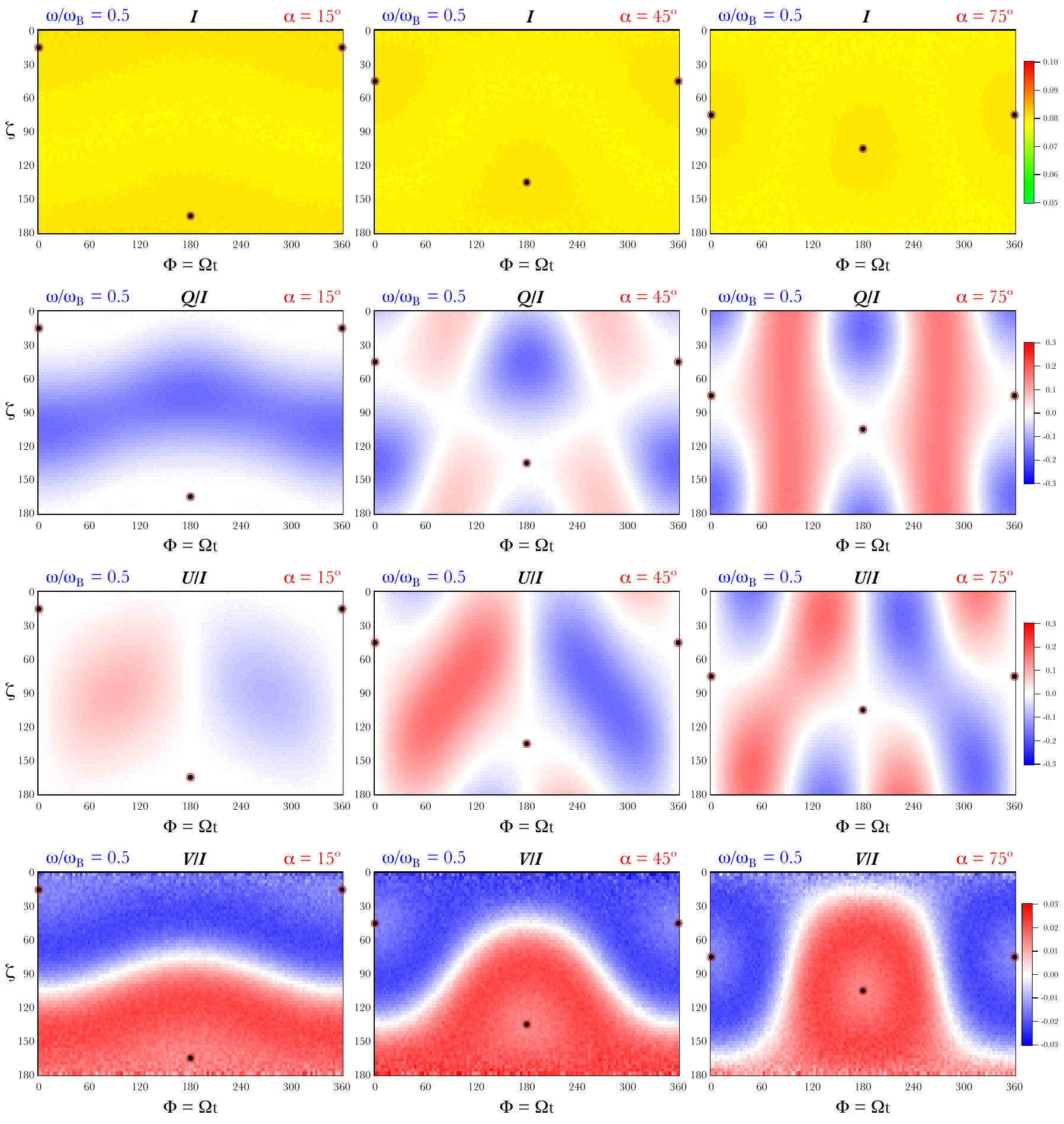}}
    \caption{
    Same as for Fig.~\ref{fig:skymap_w0-1}, but with a photon frequency ratio 
    \teq{\omega/\wcyc = 0.5} at the magnetic poles.  The intensity modulations are 
    more limited and the degree of polarization is reduced slightly relative 
    to Fig.~\ref{fig:skymap_w0-1}.  Observe that the signs of \teq{Q/I} and 
    \teq{U/I} are generally the opposite of those for the \teq{\omega/\wcyc =0.1} 
    case due to the character of the scattering cross section (see text).
    }
 \label{fig:skymap_w0-5}
\end{figure}

Fig.~\ref{fig:skymap_w0-5} presents the results for frequency ratio
\teq{\omega/\wcyc=0.5} in the LIF at the magnetic poles, chosen to closely represent the
surface X-ray emission from a moderately magnetized neutron star.  In this case, the
intensity and Stokes parameter maps share very similar \teq{\Phi - \zeta} patterns to
those in Fig.~\ref{fig:skymap_w0-1}. Yet the intensity at each surface atmospheric
locale in this case is more isotropic than that for the lower \teq{\omega/\wcyc}, with
the maximum intensity only around 1\% greater than the average value.  This is because
the photon cross section in the \teq{0.5<\omega/\wcyc<0.9} range is less sensitive to
the polarization states and scattering angles.  Therefore the local intensity beaming is
relatively muted, as can be seen from Fig. \ref{fig:slab_ap_iu}.  The result is that the
very low pulse fraction does not match those typically observed in isolated X-ray
pulsars, indicating that the region of their dominant emission is probably far inferior
to the entire surface. Alternatively, or in addition, the observed high pulse fraction
could be caused by the intrinsic beaming of the radiation emergent from atmospheres in
select surface locales. The linear polarization degree is weaker than that in
Fig.~\ref{fig:skymap_w0-1}, with a typical value around 0.1.  In this case the Stokes
\teq{Q/I} is mainly negative for small \teq{\alpha}, as opposed to the
\teq{\omega/\wcyc=0.1} case, where it is then mostly positive. This difference reflects
the transition of the dominant linear polarization mode from \teq{\parallel} to
\teq{\perp} when \teq{\omega/\wcyc} crosses the critical value
\teq{1/\sqrt{3}\sim0.577}: see Eq.~(\ref{eq:tau_eff_def}), Fig.~\ref{fig:slab_ap_iu} or
\cite{Barchas-2021-MNRAS} for a detailed discussion of this nuance pertaining to the
scattering cross section.  The Stokes \teq{U/I} also experiences this general switch of
sign for the \teq{\omega /\wcyc=0.5} case relative to the \teq{\omega /\wcyc=0.1} one. 
The circular polarization is similar to that in Fig.~\ref{fig:skymap_w0-1} with a
typical circular polarization degree around \teq{0.02}, much smaller than the linear
polarization degree. Again, this is the result of directional mixing of the fields even
for viewing angles above the magnetic poles.
\begin{figure}
    \centerline{
    \includegraphics[width=\textwidth]{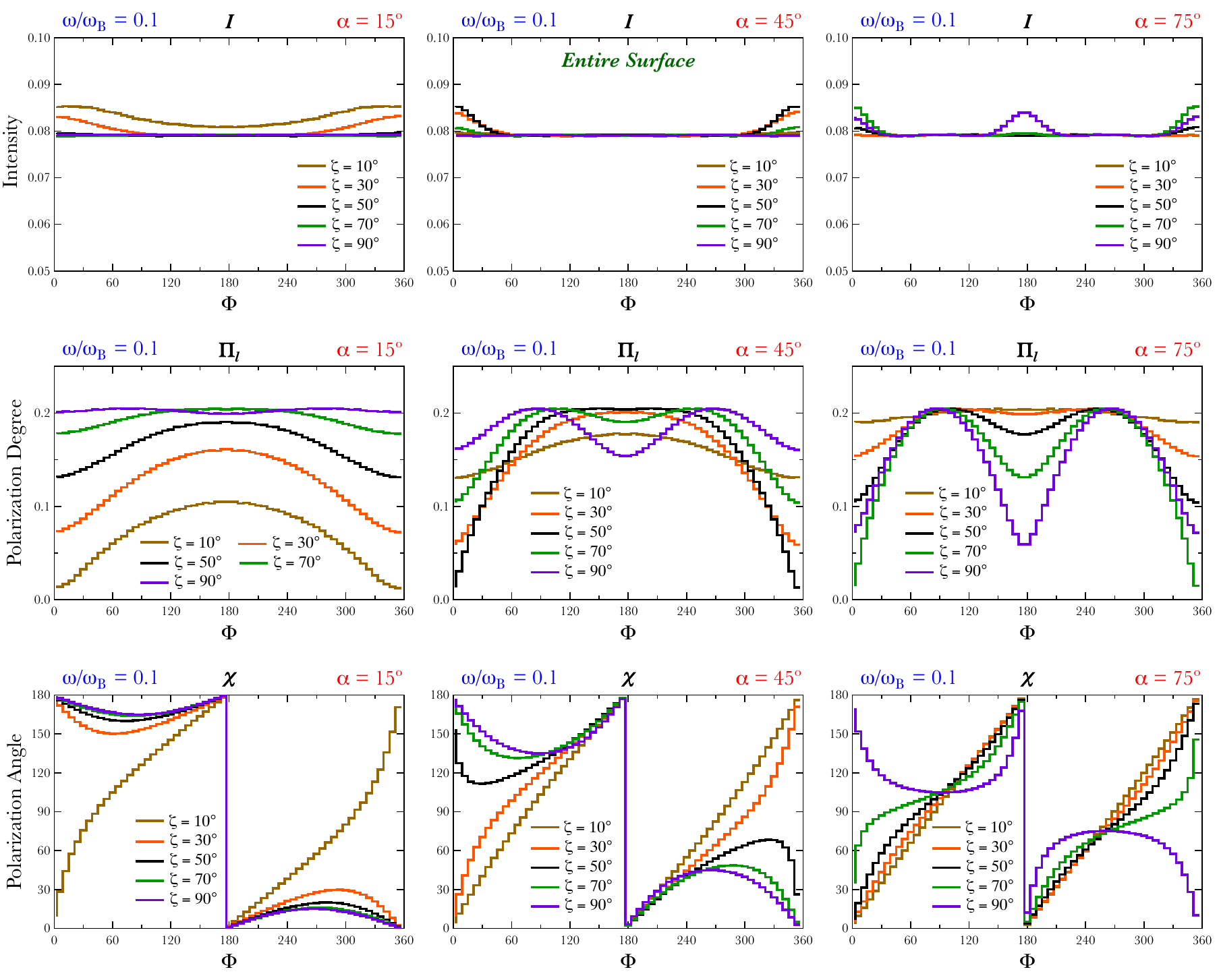}}
    \caption{
    Simulated pulse profiles for a uniformly 
    emitting entire stellar surface, for intensity \teq{I}, linear polarization degree 
    \teq{\Pi_l} and polarization angle \teq{\chi}, top to bottom, plotted as functions of the rotation phase \teq{\Phi} 
    for selected viewing angles \teq{\zeta = 10^\circ, 30^\circ, 50^\circ, 70^\circ} and \teq{90^\circ}. 
    These pulse profiles correspond to the horizontal cuts of the sky map panels in Fig.~\ref{fig:skymap_w0-1}, 
    signified in the \teq{\alpha = 45^\circ} panels by the horizontal bars that are color coded to match 
    the profiles displayed here.  Each column displays results for one of the three selected magnetic 
    inclination angles \teq{\alpha = 15^\circ , 45^\circ, 75^\circ}, and the frequency of the photons 
    is fixed so that \teq{\omega/\wcyc =0.1} at the magnetic poles in the LIF, 
    i.e.  \teq{\omega/\wcyc =0.175} at the equator.}
 \label{fig:pulse_profile_0-1_wholestar}
\end{figure}

To connect to observables pertinent to X-ray telescopes, the pulse profiles for
intensity \teq{I}, linear polarization degree \teq{\Pi_l=\sqrt{Q^2+U^2}/I} and
polarization angle \teq{\chi = [\arctan{(U/Q)}]/2} are presented in
Fig.~\ref{fig:pulse_profile_0-1_wholestar} for \teq{\omega/\wcyc = 0.1} and selected
observing angles. Pulse profiles for the \teq{\omega/\wcyc = 0.5} case are
morphologically similar, and so are not explicitly displayed in the interest of brevity.
 The profiles displayed in Fig.~\ref{fig:pulse_profile_0-1_wholestar} are derived from
horizontal sections of Fig.~\ref{fig:skymap_w0-1}, and are averaged using the symmetries
in Eq.~(\ref{eq:IQUV_sym}) so as to decrease statistical fluctuations.  The selected
observing angles are indicated as horizontal cuts in the middle column of
Fig.~\ref{fig:skymap_w0-1}.  The variation of the intensity profiles is very moderate,
as can be seen from the first row of Fig.~\ref{fig:skymap_w0-1}. The intensity maxima
are realized when the smallest instantaneous viewing angle \teq{\theta_v} to the
magnetic axis is sampled; see Eq.~(\ref{eq:cos_theta_v}).   Note that separate runs were
performed for low stellar masses and therefore essentially flat spacetime, for which the
pulse fractions (not displayed) of the intensity were somewhat higher: this just
reflected the smaller visible portions of the stellar surface in the absence of general
relativity.

The linear polarization degree \teq{\Pi_l} is an invariant under the rotation of
\teq{\hat{x}} and \teq{\hat{y}} axes in the polarization plane, thus the value of
\teq{\Pi_l} summed over the entire surface only depends on the instantaneous viewing
angle \teq{\theta_v}. Generally, \teq{\Pi_l} values are larger when \teq{\theta_v} is
greater, and this explains the observed anti-correlation of polarization degree with
intensity, as greater atmospheric radiation beaming is sampled on average. The maximum
linear polarization degree is around 20\%, values that are realized for many phases
\teq{\Phi} for most rotator and viewing geometry, with the exception of cases where the
both \teq{\alpha} and \teq{\zeta} are small. The polarization angle \teq{\chi} is
displayed in the third row of Fig.~\ref{fig:pulse_profile_0-1_wholestar}. All the
\teq{\chi} profiles jump from \teq{180^{\circ}} to \teq{0^{\circ}} at rotational phase
\teq{\Phi=180^{\circ}}, at which point the semi-major axis of the polarization ellipse
crosses the \teq{\hat{x}} axis. Such jumps are widely encountered in pulsar polarimetry
studies \citep[e.g.][]{Harding-2017-ApJ}.  In our present construction, the coordinate
axes at infinity establish \teq{U=0} at the rotational phase \teq{\Phi=\pi}, and this
precipitates the accompanying rapid swing in the value of \teq{\chi}. For viewing angle
\teq{\zeta<\alpha}, the \teq{\chi} curves are monotonically increasing in each of the
\teq{0^{\circ}<\Phi<180^{\circ}} and \teq{180^{\circ}<\Phi<360^{\circ}} ranges. For
\teq{\zeta>\alpha} the \teq{\chi} curves hover around \teq{180^{\circ}} or
\teq{0^{\circ}} for the \teq{0^{\circ}<\Phi<180^{\circ}} and
\teq{180^{\circ}<\Phi<360^{\circ}} branches respectively.  This character for the
\teq{\chi} profiles, when combined with the polarization degree and intensity
variations, affords great potential for their combined use in constraining the geometry
angles \teq{\alpha} and \teq{\zeta}, as will be touched upon in Sec.~\ref{sec:context}.

\begin{figure}
    \centerline{
    \includegraphics[width=\textwidth]{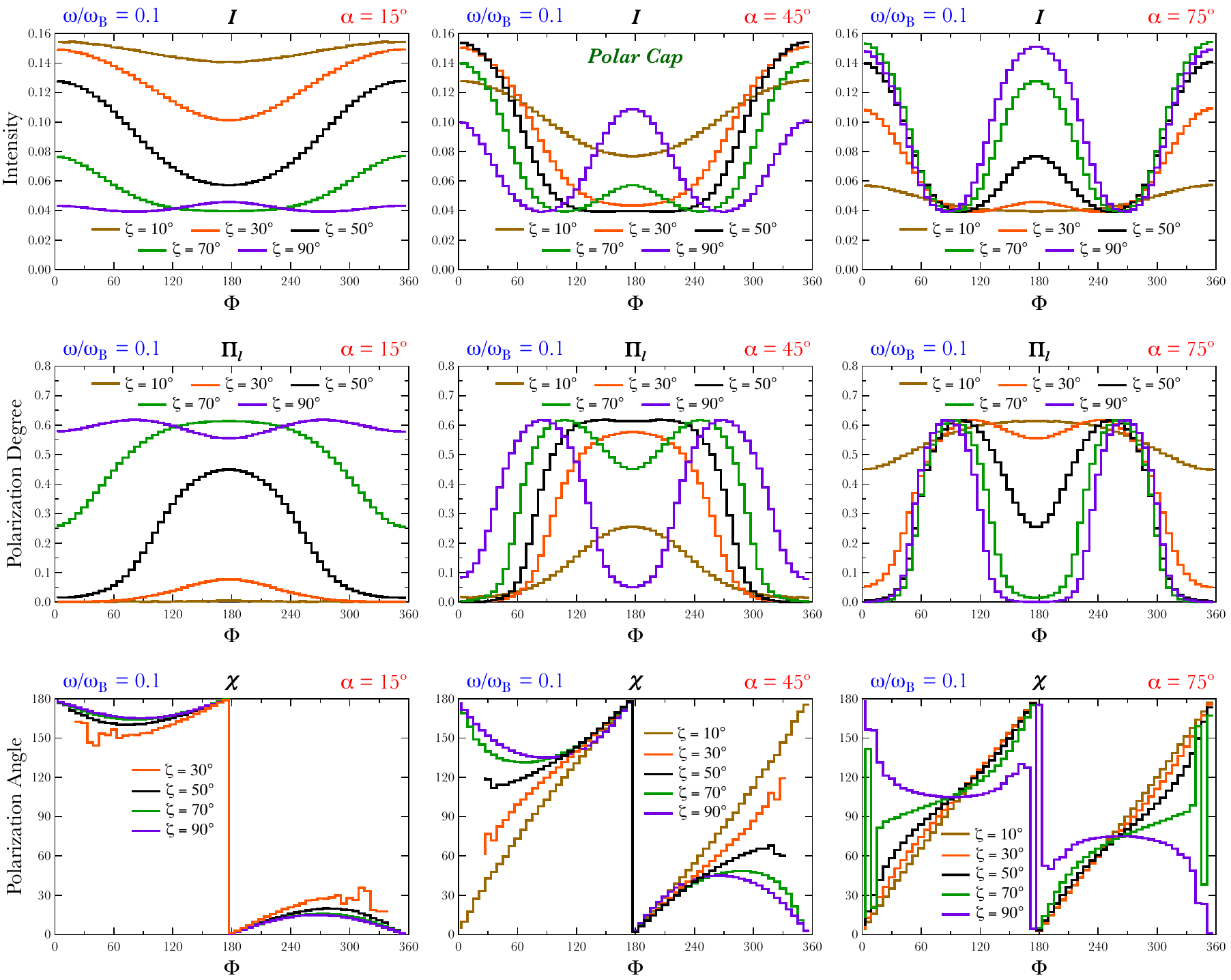}}
    \caption{
    Pulse profiles for intensity \teq{I}, linear polarization degree 
    \teq{\Pi_l} and polarization angle \teq{\chi}, top to bottom, simulated for a surface 
    with uniform emission within two antipodal polar caps spanning colatitudes 
    \teq{0^{\circ} \leq \theta \leq \theta_{\rm cap}=30^{\circ}} and 
    \teq{150^{\circ} \leq \theta \leq 180^{\circ}} degrees. 
    These are plotted as functions of the rotation phase \teq{\Phi} 
    for selected viewing angles \teq{\zeta = 10^\circ, 30^\circ, 50^\circ, 70^\circ} and \teq{90^\circ}. 
    Each column displays results for one of the three selected magnetic 
    inclination angles \teq{\alpha = 15^\circ , 45^\circ, 75^\circ}, and the frequency of the photons 
    is fixed so that \teq{\omega/\wcyc =0.1} at the magnetic poles (in the LIF).
    Observe that the \teq{\zeta} polarization angle trace for 
    \teq{\alpha = 15^{\circ}} was omitted due to poor statistics when \teq{\Pi} is low; other traces 
    are truncated, as discussed in the text. }
 \label{fig:pulse_profile_0-1_polarcap}
\end{figure}

\newpage

\subsection{Hot Polar Caps and Equatorial Belts}
 \label{sec:polar_equatorial}

A central picture emerging from Sec.~\ref{sec:entire_surface} is that if the entire
surface radiates, then the pulse profiles exhibit only very small intensity variations,
of the order of 10\% or less.  This is incompatible with numerous X-ray pulsar and
magnetar observations, likely indicating that in such cases only moderate or small
portions of the stellar surface dominate the contributions to the soft X-ray signals. 
To address this, two select cases will be considered here, one with two symmetric polar
caps where the radiative surface uniformly samples magnetic colatitudes in the range
\teq{0^{\circ} \leq \theta \leq \theta_{\rm cap}\equiv 30^{\circ}} and \teq{150^{\circ}
\leq \theta \leq 180^{\circ}}, and the other with an equatorial belt, where the active
surface region uniformly spans the colatitudes \teq{60^{\circ} \leq \theta \leq
120^{\circ}}. Polar caps are widely used for isolated neutron stars in the literature,
notably in the offerings of
\cite{Perna-2008-ApJ,Gotthelf10,Albano-2010-ApJ,Bernardini-2011-MNRAS,Younes-2020-ApJ};
see the Introduction.  Such regions are widely expected to be hotter on average due to
the much higher thermal conductivities along magnetic fields -- polar regions with
vertical fields are more likely to conduct heat to atmospheres from thermal reservoirs
deeper in the crust.  Such is a strong motivation for adopting pole-centric surface
temperature profiles with \teq{T} declining with colatitude as in
\cite{Greenstein-1983-ApJ} and numerous subsequent works. Yet, the possibility that
toroidal field components may exist near \citep[e.g.,][for magnetospheric
twists]{TLK-2002-ApJ} or in the stellar surface \citep[e.g.][for sub-surface
magneto-thermal models]{Vigano13} in magnetars promotes prospects for heat transport
away from polar zones to quasi-equatorial regions. Accordingly, equatorial belts are
treated here to provide a contrasting case. Intuitively one expects these cases to
exhibit stronger phase-dependent polarization signatures and intensity variations than
for the uniform surface, and this is borne out in the simulation runs.

Fig.~\ref{fig:pulse_profile_0-1_polarcap} displays the pulse profiles for intensity
\teq{I}, linear polarization degree \teq{\Pi_l} and polarization angle \teq{\chi} from a
star with uniformly emitting polar caps (\teq{0^{\circ} \leq \theta \leq 30^{\circ}} and
\teq{150^{\circ} \leq \theta \leq 180^{\circ}}). The intensity variation is strongly
enhanced with a typical pulsed fraction \teq{\lesssim 50\%}, contrasting the \teq{10\%}
variation from the entire surface radiating case in
Fig.~\ref{fig:pulse_profile_0-1_wholestar}. The pulse profiles of the linear
polarization degree \teq{\Pi_l} for the polar cap case are similar to those presented in
Fig.~\ref{fig:pulse_profile_0-1_wholestar}. Yet, the maximum \teq{\Pi_l} increases to
around 60\% when the instantaneous viewing angle \teq{\theta_v} to the \teq{\muvechat}
vector is close to 90\teq{^\circ}, when on average, the observer views at large angles
to the field directions at surface locales both proximate to the pole and near the
equator. This large amplitude linear polarization is also present in previous polar cap
models such as in the magnetar study of \cite{Adelsberg09}. The origin of the high
\teq{\Pi_l} for polar caps that possess a constrained field morphology is in the
reduction of the depolarization that naturally occurs when sampling a broad range of
magnetic field directions from an entire surface.  Such high polarizations are of great
interest to future X-ray polarimetry observations and can be used to diagnose the
geometric parameters (\teq{\alpha, \zeta}) of the star. The polarization angle
\teq{\chi} profiles for the polar cap case are similar to those displayed in
Fig.~\ref{fig:pulse_profile_0-1_wholestar}. For some viewing directions, small
\teq{\Pi_l} makes the \teq{\chi} profiles statistically noisy. These \teq{\chi} profiles
are truncated for the \teq{(\alpha = 45^\circ,\zeta=30^\circ, 50^\circ)} cases, and
omitted for \teq{\alpha = 15^\circ}, \teq{\zeta=10^\circ}, to improve the visualization.

While not explicitly displayed in Fig.~\ref{fig:pulse_profile_0-1_polarcap}, nor in
Fig.~\ref{fig:pulse_profile_0-1_wholestar}, there is an inherent
\teq{\alpha\leftrightarrow\zeta} interchange symmetry in the intensity and \teq{\Pi_l}
pulse profiles that can be deduced by inspection of different panels in
the upper two rows.  This is manifested in the viewing angle relation in
Eq.~(\ref{eq:cos_theta_v}), and its origin is in the uniform sampling of magnetic
azimuths on the surface.  This symmetry can be appreciated by considering
the \teq{\Phi=0} meridional plane where \teq{\Omegavec}, \teq{\muvechat} and
\teq{\kinftyvechat} are coplanar; then there are two equivalent configurations with
\teq{\muvechat} and \teq{\kinftyvechat} swapped, essentially mirror images of each
other.  These are equivalent in terms of all the local magnetic orientations sampled by
the curved photon trajectories that connect from the atmosphere to an observer at
infinity. This \teq{\alpha\leftrightarrow\zeta} symmetry does not extend to the
individual Stokes \teq{Q} and \teq{U} parameters, and therefore also not to the
polarization angle \teq{\chi}, as can be inferred from the bottom row of
Fig.~\ref{fig:pulse_profile_0-1_polarcap}. For non-uniformity of the surface in magnetic
azimuth, the degeneracy underlying this \teq{\alpha\leftrightarrow\zeta} symmetry is
broken, and furthermore, significant distortion of the pulse profiles presented in the
various figures will emerge.

\begin{figure}
    \centerline{
    \includegraphics[width=\textwidth]{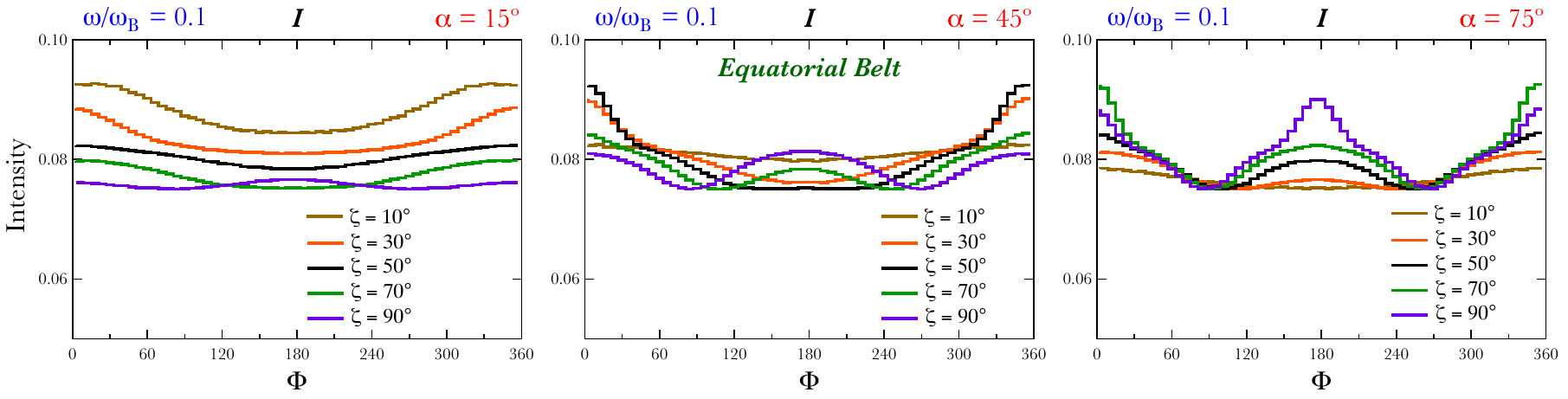}}
    \vspace{-5pt}
    \caption{
    Simulated pulse profiles for intensity \teq{I} for an equatorial 
    belt spanning colatitudes \teq{60^{\circ} \leq \theta \leq 120^{\circ}}. 
    For this case, the \teq{\Pi_l} and \teq{\chi} profiles are very similar 
    to those for the full surfaces in Fig.~6, 
    and so are not displayed here.   }
 \label{fig:pulse_profile_0-1_equatorial}
\end{figure}

To contrast the polar cap example,  Fig.~\ref{fig:pulse_profile_0-1_equatorial} presents
the intensity pulse profiles for the case where photons are emitted from an equatorial
belt in the range \teq{60^{\circ} \leq \theta \leq 120^{\circ}}, covering half of the
whole stellar surface. The intensity profiles demonstrate modest variation, around 10\%
- 20\%, marginally higher than for the entire surface emission example; this is not
surprising as the equatorial belt represents half of the stellar surface.  The
\teq{\Pi_l} and \teq{\chi} traces for the equatorial belt are very similar to those in
Fig.~\ref{fig:pulse_profile_0-1_wholestar}, with a maximum \teq{\Pi_l} around 50\%;
accordingly they are not displayed. The low pulse fraction of  the equatorial belt is a
direct consequence of the low-beaming intensity profiles at the corresponding
colatitudes, as apparent in the atmosphere results presented in
Fig.~\ref{fig:slab_ap_iu}.  In cases where soft X-ray observations of isolated neutron
stars and magnetars exhibit much higher pulse fractions, this property of the atmosphere
models indicates that polar cap locales for the origin of their signals below a few keV
are likely preferred.

\section{Discussion}
 \label{sec:context}

A signature deliverable of the {\sl MAGTHOMSCATT} code is its ability to accurately
describe the anisotropies of radiative transfer for arbitrary field directions relative
to the local zenith, and to do so in a computationally efficient manner using the high
opacity injection protocol summarized in Section~\ref{sec:injection}.  Addressing
non-vertical field directions is a non-trivial issue.  The Monte Carlo approach
accurately captures the inherent azimuthal asymmetry of photon escape around the {\bf B}
direction within the \teq{\tau \lesssim 1} moderate opacity layer in the outer
atmospheric slab.  This asymmetry is intricately intertwined with pre- and
post-scattering polarizations and photon directions, and strongly influences the values
of the Stokes parameters. The degree of intensity beaming around {\bf B} is also
precisely modeled with the Monte Carlo technique, with Fig.~\ref{fig:slab_ap_iu}
revealing prominent radiation collimation around the zenith at the magnetic poles when
\teq{\omega\ll \wcyc} that is not replicated at mid-latitude or equatorial locales. 
Such variations in local surface anisotropy profoundly impact the resultant intensity
pulse profiles for rotating neutron stars.

The pulse profiles and sky maps presented in Section~\ref{sec:extended_atmos} are for
monoenergetic photons, a restriction enabling clear discernment of principal
characteristics. These plots demonstrate that if the entire star emits equally at all
points on its surface, then the pulse fraction is generally small, as expected, and the
phase-resolved linear polarization degree is modest, typically in the \teq{5-20}\%
range. Both increase slightly when uniform emission in equatorial belts is explored (see
Fig.~\ref{fig:pulse_profile_0-1_equatorial}), and are most pronounced when polar caps
are modeled in Fig.~\ref{fig:pulse_profile_0-1_polarcap}.  Specifically, for the uniform
polar cap \teq{\omega/\wcyc =0.1} illustration, pulse fractions of \teq{30-60}\% and
linear polarization degrees of \teq{\Pi_l \sim 20-60}\% can be realized.  These both
decline somewhat when the frequency is increased to near and above the cyclotron
frequency, though such behavior is not displayed in the various figures.  All these
characteristics are consequences of the fact that the magnetic Thomson cross section is
more effective at polarizing and anisotropizing the radiation at \teq{\omega\ll\wcyc}
frequencies than at \teq{\omega > \wcyc} frequencies, where the influence of the
magnetic field is modest or small.

\subsection{Soft X-ray Signatures for a Thermal Emission Case}
 \label{sec:thermal}

In order for results from the {\sl MAGTHOMSCATT} simulation to set the scene for future
comparison with observed intensity and polarization pulse profiles, it is instructive to
illustrate their characteristics by performing simulations in certain energy bands,
encapsulating convolutions of monochromatic results like those presented in
Secs.~\ref{sec:entire_surface} and~\ref{sec:polar_equatorial}. Here we consider the
thermal emission from antipodal \teq{30^{\circ}} polar caps of a neutron star, mirroring
the geometry of Fig.~\ref{fig:pulse_profile_0-1_polarcap}, employing the standard
parameters \teq{M = 1.44M_{\odot}} and \teq{\rns = 10^6} cm used throughout this paper.
A Planck spectrum was assumed with a temperature satisfying \teq{kT = 0.42} keV as
measured by an observer at infinity. The polar caps were isothermal, so that the photon
emission was uniformly sampled across the two caps. The polar field strength was
\teq{B_p = 6.55 \times 10^{10}}Gauss, as measured by an observer at infinity.
Accordingly, in the local inertial frame at the stellar surface, the polar field was
\teq{B = 9.74 \times 10^{10}} G (\teq{\hbar\wcyc=1.13} keV), and the temperature
satisfied \teq{kT = 0.55} keV. At the rims of the polar caps, the field strength in the
LIF was \teq{B = 8.88 \times 10^{10}}Gauss (\teq{\hbar\wcyc=1.03} keV). These parameters
approximately generate the \teq{\omega /\wcyc} values addressed in prior sections of the
paper, and are suited to the pulse timing and spectral characteristics of the central
compact object PSR J0821-4300 \citep{Gotthelf10,Gotthelf13}.

The thermal simulations treated two separate energy bands sampled by an observer at
infinity, \teq{0.1- 0.5}keV and \teq{1-5}keV.  For each of these, the LIF energy of each
of the \teq{2.5 \times 10^9} photons was uniformly sampled. A multiplicative weighting
factor proportional to the Planck function was assigned to each photon according to its
energy, and this factor was applied when recording the intensity and polarization
information. The pulse profiles for intensity \teq{I} and linear polarization degree
\teq{\Pi_l} for the simulation with photons in the \teq{1-5}keV window are displayed in
Fig.~\ref{fig:pulse_profile_T_0.42keV_polarcap}. The focus on this energy range was
influenced by the fact that pulsars in the Galactic plane can be heavily absorbed at
lower X-ray energies. In this case, \teq{1.17 \leq\omega/\wcyc \leq 6.3} in the LIF at
all points on the two polar caps. The modulations of the intensity pulse profiles and
linear polarization degrees are morphologically similar to the monochromatic case in
Fig.~\ref{fig:pulse_profile_0-1_polarcap}. Yet, the intensity variation is diminished
and the linear polarization degree \teq{\Pi_l} is reduced to \teq{\lesssim 20 \%}, due
to the general character of the cross section near and above the cyclotron resonance:
see \cite{Barchas-2021-MNRAS}. The pulsed fraction for intensity from our simulations
varies from 0 to 17\%, which is comparable to the 11\% value measured in the
\teq{1.5-4.5}keV band for PSR J0821-4300 \citep{Gotthelf10}.  Moreover, the pulse shape
is similar that for this pulsar, as displayed in \cite{Gotthelf10,Gotthelf13},
suggesting good prospects for employing our simulation to constrain stellar geometry
parameters along the lines of several studies mentioned in the Introduction. The
intensity and linear polarization degree results for the \teq{0.1-0.5}keV energy band
are very similar to those exhibited in Fig.~\ref{fig:pulse_profile_T_0.42keV_polarcap},
in both pulse shape and magnitudes, are therefore are not displayed explicitly. 

\begin{figure}
    \centerline{
    \includegraphics[width=\textwidth]{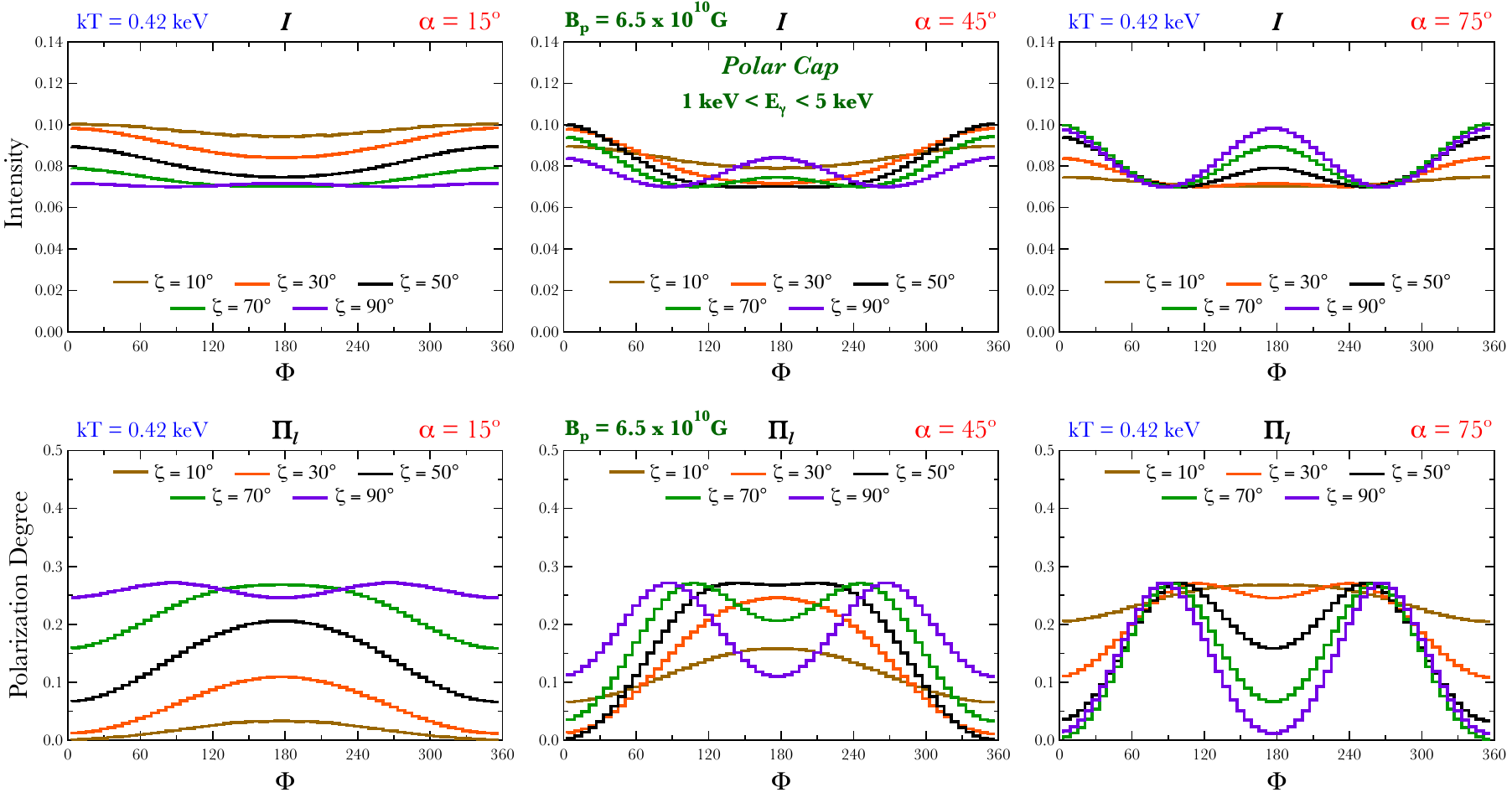}}
    \caption{
    Frequency-integrated pulse phase profiles for intensity \teq{I} (top) and linear polarization degree 
    \teq{\Pi_l} (bottom) for a uniformly emitting surface that consists of two antipodal 
    polar caps spanning \teq{30^{\circ}} in colatitude as in Fig.~\ref{fig:pulse_profile_0-1_polarcap}. 
    These are generated for a surface polar field of \teq{B_p = 6.5\times 10^{10}}Gauss
    (i.e., cyclotron energy \teq{\approx 0.75}keV) and a Planck spectrum 
    at temperature satisfying \teq{kT = 0.42}keV, both as perceived by an observer at infinity; 
    the simulation distributed photons across the energy band \teq{1-5}keV -- see text.
    The profiles are plotted for selected viewing angles 
    \teq{\zeta = 10^\circ, 30^\circ, 50^\circ, 70^\circ} and \teq{90^\circ} to the spin axis. 
    Each column displays results for one of the three selected magnetic 
    inclination angles \teq{\alpha = 15^\circ , 45^\circ, 75^\circ}. }
 \label{fig:pulse_profile_T_0.42keV_polarcap}
\end{figure}

When source counts are limited, phase-integrated linear polarizations are generally the
only obtainable polarization measures. In Table~\ref{table:linear_pol_deg}, we list two
kinds of phase-averaged linear polarization degree information from our simulations for
the different inclination and viewing angles depicted in
Fig.~\ref{fig:pulse_profile_0-1_polarcap}. First, the phase-averaged linear polarization
degree \teq{\Pi^{\rm ave}_l = \sqrt{\langle Q\rangle ^2 + \langle U\rangle^2} /\langle
I\rangle} is calculated through accumulating the Stokes \teq{Q}, \teq{U} and \teq{I} for
all recorded photons regardless of their rotational phase \teq{\Phi}.  The average
\teq{\langle I\rangle} is defined in Eq,~(\ref{eq:Pi_l_ave}), with similar definitions
for \teq{\langle Q\rangle} and \teq{\langle U\rangle}. The \teq{ \Pi^{\rm ave}_l }
values are around a few percent for most viewing directions, which are smaller than the
typical phase-resolved linear polarization degrees presented in
Fig~\ref{fig:pulse_profile_T_0.42keV_polarcap}. This is because the phase-resolved
polarization angle \teq{\chi} is strongly dependent on \teq{\Phi}, engendering strong
cancellation in \teq{Q/I} and \teq{U/I} measures for different photons, which is
signified by the alternating blue and red regions of the maps in
Figs.~\ref{fig:skymap_w0-1} and~\ref{fig:skymap_w0-5}. It is noticeable that \teq{
\Pi^{\rm ave}_l } increases monotonically with increasing \teq{\zeta}, a correlation
that is driven by the observer generally viewing more obliquely to the surface field
direction on average.  When \teq{\zeta} is small, the viewer sees a polarization angle
that rotates around the line of sight so that strong cancellation in both \teq{Q} and
\teq{U} results when averaging over pulse phase.

As an alternative measure, the average of the phase-resolved linear polarization degree
\teq{ \overline{\Pi}^{\rm pr}_l } is defined as the average of the phase-dependent
\teq{\Pi_l} values weighted by the intensity pulse profiles:
\begin{equation}  
       \overline{\Pi}^{\rm pr}_l  \; \equiv\; \dover{1}{\langle I\rangle} 
       \int \sqrt{ \bigl[Q(\Phi )\bigr]^2 + \bigl[U(\Phi )\bigr]^2} \, \dover{d \Phi}{2\pi} 
       \quad ,\quad
       \langle I\rangle \; =\; \int I(\Phi) \, \dover{d \Phi}{2\pi}  \quad.
 \label{eq:Pi_l_ave}
\end{equation}
The average \teq{ \overline{\Pi}^{\rm pr}_l } gives the typical value of the phase-dependent linear polarization degree,
and is in the range of \teq{1-25}\% for the cases listed in Table~\ref{table:linear_pol_deg}. 
Since there is no cancellation in the averaging process, \teq{ \overline{\Pi}^{\rm pr}_l } is always greater than \teq{ \Pi^{\rm ave}_l}.  Observe that while  \teq{ \overline{\Pi}^{\rm pr}_l } increases 
monotonically with \teq{\zeta} when \teq{\alpha \leq 45^{\circ}}, it actually declines as \teq{\zeta}
increases for the \teq{\alpha = 75^{\circ}} simulations.  This is a consequence of 
the fact that for large \teq{\alpha}, perspectives with smaller \teq{\zeta} permit the observer 
to more often view more obliquely to the surface {\bf B} direction, generating high 
\teq{\Pi (\Phi )}.  In contrast, large \teq{\zeta} for such quasi-perpendicular rotators leads to 
low values of \teq{\Pi (\Phi )} when the observer is looking almost parallel or anti-parallel 
to the local field direction, lowering \teq{ \overline{\Pi}^{\rm pr}_l } accordingly. 
The fact that both polarization averages are higher for the 
\teq{1-5}keV band than for the \teq{0.1-0.5}keV band reflects a stronger partial cancellation of 
Stokes parameters in the sub-cyclotronic domain.  With the launch of the IXPE mission this last December, 
phase-resolved polarization measurements, \teq{\Pi_l(\Phi )}, of bright isolated neutron stars such as 
magnetars, will soon become available.  Added to this library will be phase-averaged 
characteristics for relatively faint sources. This will allow us to confront our theoretical predictions with
observationally determined polarization properties, thereby providing potential 
for discerning luminous atmospheric locations and neutron star geometry parameters in detail.

\begin{deluxetable}{c c  c|cc|cc|cc|cc|cc}[htb!]
\centerwidetable
\movetabledown = 50mm
\tablecolumns{13}

\tablecaption{Thermal Emission Simulations: Phase-averaged Linear
Polarization Degrees}
\label{table:linear_pol_deg}

\tablehead{ \colhead{} & \multicolumn{4}{c|}{\teq{\alpha = 15^{\circ}}}  & \multicolumn{4}{c|}{\teq{\alpha = 45^{\circ}}} & \multicolumn{4}{c}{\teq{\alpha = 75^{\circ}}} \\ 
\hline
\colhead{} & \multicolumn{2}{c}{0.1-0.5 keV} & \multicolumn{2}{c|}{1-5 keV} & \multicolumn{2}{c}{0.1-0.5 keV} & \multicolumn{2}{c|}{1-5 keV} & \multicolumn{2}{c}{0.1-0.5 keV} & \multicolumn{2}{c}{1-5 keV}\\
\hline
\colhead{} & \multicolumn{2}{c|}{\teq{ \Pi^{\rm ave}_l   \qquad  \overline{\Pi}^{\rm pr}_l }  } & \multicolumn{2}{c|}{\teq{ \Pi^{\rm ave}_l   \qquad  \overline{\Pi}^{\rm pr}_l }  } & \multicolumn{2}{c|}{\teq{ \Pi^{\rm ave}_l   \qquad  \overline{\Pi}^{\rm pr}_l }  } & \multicolumn{2}{c|}{\teq{ \Pi^{\rm ave}_l   \qquad  \overline{\Pi}^{\rm pr}_l }  } & \multicolumn{2}{c|}{\teq{ \Pi^{\rm ave}_l   \qquad  \overline{\Pi}^{\rm pr}_l }  } & \multicolumn{2}{c}{\teq{ \Pi^{\rm ave}_l   \qquad  \overline{\Pi}^{\rm pr}_l }  }   }
\startdata
\teq{\zeta = 10^{\circ}} & 0.0030 & 0.0095 &0.0051 &0.0168 & 0.0021 & 0.0810 & 0.0019 & 0.1112 & 0.0026 & 0.1866 & 0.0032 & 0.2409\\
\teq{\zeta = 30^{\circ}} &0.0302 & 0.0384 &0.0450 &0.0581 & 0.0123 & 0.0936 &0.0152 & 0.1264 & 0.0202 & 0.1599 & 0.0257 & 0.2099\\
\teq{\zeta = 50^{\circ}} &0.0907 & 0.1021 &0.1202 &0.1355 & 0.0255 & 0.1104 & 0.1264 & 0.1509 & 0.0420 & 0.1252 & 0.0565 & 0.1662\\
\teq{\zeta = 70^{\circ}} &0.1571 & 0.1705 &0.2029 &0.2199 & 0.0398 & 0.1300 & 0.0529 & 0.1732 & 0.0599 & 0.0966 & 0.0806 & 0.1316\\
\teq{\zeta = 90^{\circ}} &0.1839 & 0.1975 &0.2404  &0.2583 & 0.0461 & 0.1393 & 0.0603 & 0.1827 & 0.0661 & 0.0861 & 0.0896 & 0.1187
\enddata
\end{deluxetable}
\vspace{-30pt}

\subsection{Interpreting the Sky Maps: Stellar Geometry Diagnostics}

The pattern structure of the sky maps serves as a fingerprint of the neutron star rotator 
geometry.  The general appearance is governed by the loci of the zeros for the Stokes 
parameters that partition zones of opposite sign.  The specification of these loci can
be understood as follows.  Due to the axial symmetry of the magnetic field configuration,
the Stokes \teq{Q} equals zero when the projection of the magnetic axis \teq{\muvechat}
on the observer's sky makes a 45\teq{^\circ} or 135\teq{^\circ}
angle with the projection of the rotation axis \teq{\Omegavec} on the sky plane.
The criteria can be expressed as \teq{\muvechat\cdot\hat{x}/\sin{\theta_v}=\pm 1/\sqrt{2}},
where \teq{\hat{x}} and \teq{\theta_v} are defined in Eq.~(\ref{eq:x_y_obs_polarize})
and Eq.~(\ref{eq:cos_theta_v}).   This relation can be simplified to
\begin{equation}
  \sin{\alpha}\cos{\Phi}\cos{\zeta} - \cos{\alpha}\sin{\zeta}
  \;=\; \pm\sin{\alpha}\sin{\Phi} \quad ,
\label{eq:Q_skymap_zero}
\end{equation}
where the \teq{\pm} sign on the right identifies the \teq{\Phi \to 2\pi - \Phi} symmetry 
of the geometry.  As the star rotates, these criteria sample different \teq{\zeta} angles
and trace two branches of white zero-value paths in the \teq{\Phi-\zeta} maps.
Similarly, Stokes \teq{U} equals zero when the projection of the magnetic axis on 
the observer's sky is parallel or orthogonal to the projection of the rotation axis on 
the sky, which yields \teq{\muvechat\cdot\hat{x}/\sin{\theta_v} =\pm 1}.  This defines 
the morphology for the zero-value loci in the \teq{U} maps:
\begin{equation}
  \sin{\Phi}\;=\;0 \quad\text{or} \quad\tan{\zeta}
  \;=\;\tan{\alpha}\cos{\Phi} \quad ,
\label{eq:U_skymap_zero}
\end{equation}
with both relations encompassing \teq{\Phi \to 2\pi - \Phi} symmetry. Finally, the loci
for \teq{V=0} constitute viewing directions where regions of opposite polarity are
sampled equally across the stellar surface so that their emergent circular polarization
signals precisely cancel. The only observer directions that satisfy this criterion for
uniform surface illumination are those above the magnetic equator, so that the \teq{V=0}
loci in Figs.~\ref{fig:skymap_w0-1} and~\ref{fig:skymap_w0-5} satisfy \teq{\muvechat
\cdot \kinftyvechat = 0}, i.e. \teq{\theta_v=90^{\circ}}.

Observe that since these loci for zeros of the Stokes parameters depend only on the
geometrical configuration of the three vectors \teq{\Omegavec}, \teq{\kinftyvechat} and
the precessing \teq{\muvechat}, they are independent of photon frequency, a property
that is apparent when comparing Figs.~\ref{fig:skymap_w0-1} and~\ref{fig:skymap_w0-5};
they are also independent of stellar mass and thus spacetime curvature. In principle,
the determination of the phases for zeros of the Stokes Q and U parameters (or,
equivalently, \teq{\Pi_l} and \teq{\chi}) in concert with the intensity pulse profile
provides powerful diagnostics on the stellar geometry parameters. Yet note that
departures from azimuthal independence for the surface emission will introduce
distortions to the shapes of these loci of zeros, thereby complicating the stellar
parameter determination.

\newpage

\section{Conclusion}
 \label{sec:conclusion}

In this paper, we present our Monte Carlo simulation {\sl MAGTHOMSCATT} that models
radiation transport and scattering in atmospheric slabs, and generates intensity and
polarization signatures from extended surface regions of magnetized neutron stars. The
code captures the azimuthally asymmetric beaming and polarization information for each
magnetic field orientation and generates pulse profiles for intensity and polarization
signatures from extended atmospheres. This simulation of the magnetic Thomson scattering
in high opacity environs is versatile, and can also be deployed in neutron star settings
remote from their surfaces, for example to treat the accretion columns of X-ray pulsars
and the transfer of radiation in bursts in magnetar magnetospheres. The code also
implements general relativistic ray tracing and parallel transport of the Stokes
parameters. These GR effects alter the observed intensity and polarization signals,
changes that are apparent in \teq{I, Q, U, V} traces that are resolved in rotational
phase; they therefore influence the determination of the stellar geometry parameters
using phase-resolved observations. For a uniformly emitting star with \teq{M =
1.44M_{\odot}}, \teq{\rns = 10^6} cm and \teq{\omega/\wcyc=0.1}, the effects of general
relativity reduce the intensity variation from 30\% to 6\% (see
Fig.~\ref{fig:pulse_profile_0-1_wholestar}), and change the general viewing direction
for a maximum of intensity from \teq{\theta_v = 90^{\circ}} in flat spacetime (roughly
over the magnetic equator) to \teq{\theta_v = 0^{\circ}} in GR, i.e. over the magnetic
poles (see Fig.~\ref{fig:pulse_profile_0-1_wholestar}). We find that for a neutron star
with uniformly emitting polar caps spanning a half angle \teq{30^\circ} in magnetic
colatitude, the phase-resolved linear polarization degree can be as much as 60\% for
monochromatic photons.

To provide context for soft X-ray observations, we then simulated intensity and
polarization signatures from the polar cap regions of a neutron star with parameters
comparable to the central compact object PSR J0821-4300. This element considered photons
from a Planck spectrum and assumed isothermal conditions across the polar caps. Photons
in two energy bands were simulated, and the average linear polarization information was
listed in Table~\ref{table:linear_pol_deg}. For this specific case, the phase-resolved
linear polarization degree is generally in the range 10-25\% when the viewing angle
\teq{\zeta} is not too small, though the phase-averaged polarization is noticeably
smaller, at the 1-20\% level.  Accordingly, it is clear that the phase-dependent
polarization signatures from these simulations provide better prospects for exploitation
using sensitive soft X-ray polarimeters in probing stellar geometry parameters. Future
extensions of this work will focus on introducing hydrostatic structure to the
atmosphere model. This will determine the temperature stratification in the outermost
layers of the star and therefore define the interplay between polarization transport and
photon frequencies that sample different physical depths within atmospheres. Free-free
opacity, important at photon energies below around 2-3 keV, will also be incorporated,
thereby modifying the transport from a purely magnetic Thomson one. In addition, the
influences of vacuum birefringence on polarization transfer in twisted magnetospheres
will be addressed.  With these enhancements, our atmospheric radiative transfer code
will be suited to addressing neutron star observations acquired by IXPE and future
advanced X-ray polarimeters.

\vspace{-10pt}
\acknowledgments

We thank the referee for suggestions helpful to improving the presentation.
M.~G.~B. acknowledges the support of the National Science Foundation
through grant AST-1813649.  G.~Y. acknowledges the support of 
the NASA Postdoctoral Program at the Goddard Space Flight Center, 
administered by the USRA through a contract with NASA.

\bibliographystyle{aasjournal} 

\vspace{-10pt}
\begin{thebibliography}{99}

\bibitem[\protect\citeauthoryear{Abarr et al.}{2020}]{Abarr-2020-ApJ} 
Abarr Q., Baring M., Beheshtipour B., Beilicke M., de Geronimo G., Dowkontt P., Errando M., et al., 2020, ApJ, 891, 70. doi:10.3847/1538-4357/ab672c

\bibitem[Albano et al.(2010)]{Albano-2010-ApJ}
Albano, A., Turolla, R., Israel, G.~L., et al.\ 2010, \apj, 722, 788. doi:10.1088/0004-637X/722/1/788

\bibitem[Barchas(2017)]{Barchas17}
Barchas, J.~A.\ 2017, PhD Thesis, Rice University.

\bibitem[Barchas, Hu \& Baring(2021)]{Barchas-2021-MNRAS}
Barchas, J.~A., Hu, K. \& Baring, M.~G. 2021, \mnras, 500, 5369. doi:10.1093/mnras/staa3541  

\bibitem[Beloborodov(2002)]{Beloborodov-2002-ApJ} 
Beloborodov, A.~M.\ 2002, \apjl, 566, L85. doi:10.1086/339511

\bibitem[Bernardini et al.(2011)]{Bernardini-2011-MNRAS}
Bernardini, F., Perna, R., Gotthelf, E.~V., et al.\ 2011, \mnras, 418, 638. doi:10.1111/j.1365-2966.2011.19513.x

\bibitem[Braje et al.(2000)]{Braje-2000-ApJ} 
Braje, T.~M., Romani, R.~W., \& Rauch, K.~P.\ 2000, \apj, 531, 447. doi:10.1086/308448

\bibitem[Chandrasekhar(1960)]{Chandra60}
Chandrasekhar, S.\ 1960, {\it Radiative Transfer} (Dover, New York)

\bibitem[Chou(1986)]{Chou-1986-ApSS}
Chou C.~K., 1986, ApSS, \vol{121}{333} doi:10.1007/BF00653705

\bibitem[Fern{\'a}ndez \& Thompson(2007)]{Fernandez07}
Fern{\'a}ndez, R., \& Thompson, C.\ 2007, ApJ, 660, 615. doi:10.1086/511810

\bibitem[Fern{\'a}ndez \& Davis(2011)]{Fernandez11}
Fern{\'a}ndez, R., \& Davis, S.~W.\ 2011, ApJ, 730, 131. doi:10.1088/0004-637X/730/2/131

\bibitem[Gonthier \& Harding(1994)]{GH94} 
Gonthier, P.~L. \& Harding, A.~K. 1994, ApJ, \vol{425}{767} doi:10.1086/174020

\bibitem[Gotthelf et al.(2010)]{Gotthelf10}
Gotthelf, E.~V., Perna, R., \& Halpern, J.~P.\ 2010, ApJ, \vol{724}{1316}.  doi: 10.1088/0004-637X/724/2/1316

\bibitem[Gotthelf et al.(2013)]{Gotthelf13}
Gotthelf, E.~V., Halpern, J.~P., \& Alford, J.\ 2013, ApJ, \vol{765}{58}. doi:  10.1088/0004-637X/765/1/58

\bibitem[Greenstein \& Hartke(1983)]{Greenstein-1983-ApJ} 
Greenstein, G. \& Hartke, G.~J.\ 1983, \apj, 271, 283. doi:10.1086/161195

\bibitem[Harding \& Muslimov(1998)]{Muslimov-1998-ApJ} 
Harding, A.~K. \& Muslimov, A.~G.\ 1998, \apj, 500, 862. doi:10.1086/305763

\bibitem[Harding \&  Kalapotharakos(2015)]{Harding-2015-ApJ} 
Harding, A.~K. \& Kalapotharakos, C.\ 2015, \apj, 811, 63. doi:10.1088/0004-637X/811/1/63

\bibitem[Harding \& Kalapotharakos(2017)]{Harding-2017-ApJ} 
Harding, A.~K. \& Kalapotharakos, C.\ 2017, \apj, 840, 73. doi:10.3847/1538-4357/aa6ead

\bibitem[Heyl et al.(2003)]{Heyl-2003-MNRAS} 
Heyl, J.~S., Shaviv, N.~J., \& Lloyd, D. 2003, \mnras, 342, 134. doi: 10.1046/j.1365-8711.2003.06521.x

\bibitem[Ho \& Lai(2001)]{HoLai-2001-MNRAS}
Ho, W.~C.~G., \& Lai, D.\ 2001, \mnras, 327, 1081. doi:10.1046/j.1365-8711.2001.04801.x
   
\bibitem[Hu et al.(2019)]{Hu-2019-MNRAS} 
Hu, K., Baring, M.~G., Wadiasingh, Z., et al.\ 2019, \mnras, 486, 3327. doi:10.1093/mnras/stz995


\bibitem[Jahoda et al.(2019)]{Jahoda19} 
Jahoda, K., Krawczynski, H., Kislat, F., et al. 2019, arXiv e-prints, arXiv:1907.10190.

\bibitem[{{Lai} \& {Ho}(2003)}]{Lai2003}
Lai, D., \& Ho, W.~C.~G. 2003, Phys. Rev. Lett, 91, 071101. doi:10.1103/PhysRevLett.91.071101

\bibitem[Medin \& Lai(2006)]{Medin06}
Medin, Z., Lai, D.\ 2006, Phys. Rev. A, \vol{74}{062508} doi:10.1103/PhysRevA.74.062508 

\bibitem[Medin \& Lai(2007)]{Medin07}
Medin, Z., Lai, D.\ 2007, MNRAS, \vol{382}{1833} 10.1111/j.1365-2966.2007.12492.x 

\bibitem[{{M{\'e}sz{\'a}ros}(1992)}]{Meszaros-1992}
M{\'e}sz{\'a}ros, P. 1992, {\it High-energy Radiation from Magnetized Neutron Stars},
   (University of Chicago Press, Chicago)

\bibitem[\protect\citeauthoryear{M\'esz\'aros \& Bonazzola}{1981}]{MB81_ApJ}
M\'esz\'aros P., \& Bonazzola S., 1981, ApJ, \vol{251}{695} doi:10.1086/159515 

\bibitem[\protect\citeauthoryear{M\'esz\'aros et al.}{1988}]{Meszaros-1988-ApJ} 
M\'esz\'aros P., Novick R., Szentgyorgyi A., et al., 1988, \apj,\vol{324}{1056} doi:10.1086/165962

\bibitem[Miller et al.(2019)]{Miller-2019-ApJL}
Miller, M.~C., Lamb, F.~K., Dittmann, A.~J., et al.\ 2019, ApJL, \vol{887}{L24} doi:10.3847/2041-8213/ab50c5 

\bibitem[Misner et al.(1973)]{MTW-1973}
Misner, C.~W., Thorne, K.~S., \& Wheeler, J.~A. 1973, {\it Gravitation} (San Francisco: W.H. Freeman and Co.)
   
\bibitem[Muslimov \& Tsygan(1986)]{Muslimov86} 
Muslimov, A.~G., \& Tsygan, A.~I. 1986, Sov. Astr. \vol{30}{567} 

\bibitem[Nobili et al.(2008)]{Nobili-2008-MNRAS} 
Nobili, L., Turolla, R., \& Zane, S.\ 2008, \mnras, 386, 1527. doi:10.1111/j.1365-2966.2008.13125.x

\bibitem[Nollert et al.(1989)]{Nollert-1989-AA} 
Nollert, H.-P., Ruder, H., Herold, H., \& Kraus, U.\ 1989, A\&A, 208, 153.

\bibitem[{\"O}zel(2001)]{Ozel-2001-ApJ}
{\"O}zel, F.\ 2001, \apj, 563, 276. doi:10.1086/323851
 
\bibitem[Pavlov et al.(1994)]{Pavlov94}
Pavlov, G.~G., Shibanov, Yu.~A., Ventura, J., et al.\ 1994, A\&A, \vol{289}{837}

\bibitem[Pavlov \& Zavlin(2000)]{Pavlov-2000-ApJ} 
Pavlov, G.~G., \& Zavlin, V.~E. 2000, \apj, 529, 1011. doi: 10.1086/308313

\bibitem[Pechenick et al.(1983)]{Pechenick-1983-ApJ} 
Pechenick, K.~R., Ftaclas, C., \& Cohen, J.~M.\ 1983, \apj, 274, 846. doi:10.1086/161498

\bibitem[Perna \& Gotthelf(2008)]{Perna-2008-ApJ} 
Perna, R. \& Gotthelf, E.~V.\ 2008, \apj, 681, 522. doi:10.1086/588211

\bibitem[Pineault (1977)]{Pineault-1977-MNRAS} 
Pineault, S. 1977, \mnras,\vol{179}{691} doi: 10.1093/mnras/179.4.691

\bibitem[Potekhin et al.(2004)]{Potekhin04}
Potekhin, A.~Y., Lai, D., Chabrier, G., et al.\ 2004, ApJ, \vol{612}{1034} doi:10.1086/422679 

\bibitem[Poutanen (2020)]{Poutanen-2020-AA} 
Poutanen, J. 2020, A\&A,\vol{640}{A24} doi: 10.1051/0004-6361/202037471

\bibitem[Psaltis et al.(2014)]{Psaltis-2014-ApJ} 
Psaltis, D., {\"O}zel, F., \& Chakrabarty, D.\ 2014, \apj, 787, 136. doi:10.1088/0004-637X/787/2/136

\bibitem[Riffert \& M\'esz\'aros(1988)]{Riffert-1988-ApJ} 
Riffert, H. \& M\'esz\'aros, P.\ 1988, \apj, 325, 207. doi:10.1086/165996

\bibitem[Riley et al.(2019)]{Riley-2019-ApJL}
Riley, T.~E., Watts, A.~L., Bogdanov, S., et al.\ 2019, ApJL, \vol{887}{L21} doi:10.3847/2041-8213/ab481c

\bibitem[Rybicki \& Lightman(1979)]{Rybicki79}
Rybicki, G.~B., \& Lightman, A.~P. 1979, {\it Radiative Processes in Astrophysics} (Wiley-Interscience, New York)

\bibitem[Shibanov et al.(1992)]{Shibanov92}
Shibanov, Yu.~A., Zavlin, V.~E., Pavlov, G.~G., et al.\ 1992, A\&A, \vol{266}{313}

\bibitem[Story \& Baring(2014)]{Story-2014-ApJ} 
Story, S.~A. \& Baring, M.~G.\ 2014, \apj,\vol{790}{61} doi:10.1088/0004-637X/790/1/61

\bibitem[Suleimanov et al.(2009)]{Suleimanov09}
Suleimanov V., Potekhin A.~Y., Werner K.\ 2009, A\&A, \vol{500}{891} doi:10.1051/0004-6361/200912121

\bibitem[Taverna \& Turolla(2017)]{Taverna17}
Taverna, R. \& Turolla, R.\ 2017, \mnras,\vol{469}{3610} doi:10.1093/mnras/stx1086
   
\bibitem[Taverna et al.(2020)]{Taverna20}
Taverna, R., Turolla, R., Suleimanov, V., et al.\ 2020, \mnras,\vol{492}{5057} doi:10.1093/mnras/staa204

\bibitem[Thompson et al.(2002)]{TLK-2002-ApJ} 
Thompson, C., Lyutikov, M., \& Kulkarni, S.~R.\ 2002, \apj, \vol{574}{332} doi:10.1086/340586
   
\bibitem[van Adelsberg \& Lai(2006)]{Adelsberg06}
van Adelsberg, M., Lai, D.\ 2006, \mnras,\vol{373}{1495} doi:10.1111/j.1365-2966.2006.11098.x

\bibitem[van Adelsberg \& Perna(2009)]{Adelsberg09}
van Adelsberg, M., \& Perna, R. 2009, \mnras,\vol{399}{1523} doi:10.1111/j.1365-2966.2009.15374.x

\bibitem[Ventura, Nagel, \& M\'esz\'aros(1979)]{Ventura-1979-ApJ} 
Ventura J., Nagel W. \& M\'esz\'aros P., 1979, \apjl, \vol{233}{L125} doi:10.1086/183090 

\bibitem[Vigan{\`o} et al.(2013)]{Vigano13}
Vigan{\`o}, D., Rea, N., Pons, J.~A., et al.\ 2013, \mnras,\vol{434}{123} doi:10.1093/mnras/stt1008

\bibitem[Wasserman \& Shapiro(1983)]{WS83} 
Wasserman, I. \& Shapiro, S.~L. 1983, ApJ, \vol{265}{1036} doi:10.1086/160745

\bibitem[Weinberg(1972)]{Weinberg72}
   Weinberg, S. 1972, {\it Gravitation and Cosmology: Principles and Applications of the 
   General Theory of Relativity} (Wiley \& Sons, New York).

\bibitem[Weisskopf et al.(2016)]{Weisskopf16}
Weisskopf, M.~C., Ramsey, B., O'Dell, S., et al.\ 2016, \procspie, 9905, 990517. doi:10.1117/12.2235240

\bibitem[Whitney(1991a)]{Whitney-ApJS91}
Whitney, B.~A. 1991, ApJ Supp., 75, 1293. doi:10.1086/191560

\bibitem[Younes et al.(2020)]{Younes-2020-ApJ}
Younes, G., Baring, M.~G., Kouveliotou, C., et al.\ 2020, \apjl,\vol{889}{L27} doi:10.3847/2041-8213/ab629f

\bibitem[Zane et al.(2000)]{Zane00}
Zane S., Turolla R., Treves A.\ 2000, ApJ, \vol{537}{387} doi:10.1086/309027

\bibitem[Zhang et al.(2016)]{Zhang16}
Zhang, S.~N., Feroci, M., Santangelo, A., et al.\ 2016, in Proc. SPIE, Vol. 9905.  doi:10.1117/12.2232034

\end{thebibliography}

\begin{flushleft}
\bf{Appendix A: Magnetic Thomson Scattering Formalism}
\end{flushleft}
 
For the deployment of complex electric field vector information for photon transport in
the slab, here we briefly summarize the key elements of the scattering: see 
\cite{Barchas-2021-MNRAS} for extensive details.  The differential and total cross sections 
for a magnetic Thomson scattering are expressed in terms of the electric vectors of 
the incoming and outgoing electromagnetic waves.  If the incident electromagnetic wave 
propagates in the direction of the unit vector \teq{\kvechat_i}, the time portion of its electric 
field vector is encapsulated in \teq{\Evec(t) = \calEvec_i e^{-i\omega t}}, with 
\teq{\calEvec_i\cdot\kvechat_i=0}.  The wave's oscillating electric field accelerates the 
non-relativistic electron subject to the influence of a magnetic field \teq{\Bvec \equiv B\Bvechat}.  
The motion is described by the Newton-Lorentz equation, whose solution generates an acceleration
\begin{equation}
   \avec \; =\; - \dover{e}{m_e} \, \dover{\alphavec \, \vert \calEvec_i \vert }{\omega^2 - \wcyc^2} \, e^{-i\omega t}
   \qquad \hbox{for}\qquad
   \alphavec \; =\; \omega ^2 \calEvechat_i
      -i \omega \wcyc \,\calEvechat_i \times \Bvechat
      - \wcyc^2 ( \calEvechat_i \cdot \Bvechat ) \Bvechat \quad .
 \label{eq:scatt_accel}
\end{equation}
Here, \teq{\wcyc = eB/m_ec} is the electron cyclotron frequency, and a scaled 
incident polarization vector \teq{\calEvechat_i = \calEvec_i /\vert \calEvec_i \vert} 
has been introduced to simplify ensuing expressions for the differential and total 
cross section.  In general, \teq{\kvechat_i \times \Bvechat \neq \mathbf{0}} so that 
\teq{\calEvec_i \cdot \Bvechat \neq \mathbf{0}}.  The \teq{\calEvechat_i \times \Bvechat} 
term in \teq{\alphavec} seeds circular polarization of the emergent wave, 
its influence being maximized near the cyclotron frequency; the other two terms 
precipitate linear polarization.

The electric field vector \teq{\calEvec_f} for the scattered (final) electromagnetic wave 
can quickly be written down using the non-relativistic dipole radiation formula 
\citep[e.g.,][]{Rybicki79}.  As in \cite{Barchas-2021-MNRAS}, here this polarization 
is expressed in normalized form, \teq{\calEvechat_f = \calEvec_f /\vert \calEvec_f \vert}, 
and the differential cross section for magnetic Thomson scattering can be obtained as 
the ratio of Poynting fluxes for the final and initial waves:
\begin{equation}
   \calEvechat_f \;\equiv\; \dover{\calEvec_f}{ \vert \calEvec_i \vert} \; =\; 
    \dover{ \kvechat_f \times \bigl( \kvechat_f \times \alphavec \bigr)}{\omega^2 - \wcyc^2}
    \quad \Rightarrow \quad
  \dover{d\sigma}{d\Omega_f} 
   \;=\; r_0^2 \, \calEvechat_f \cdot \calEvechat_f^*    
   \; =\; r_0^2 \, \dover{ \bigl( \kvechat_f \times \alphavec \bigr) \cdot  
       \bigl( \kvechat_f \times \alphavec^{\ast} \bigr)}{ (\omega^2 - \wcyc^2 )^2}\quad ,
 \label{eq:dsig_magThom}
\end{equation}
noting that this used the \teq{\vert \calEvechat_i\vert =1} normalization result.  Here, 
\teq{r_0=e^2/m_ec^2} is the familiar classical electron radius.  The full expression for 
\teq{d\sigma/d\Omega_f} in terms of the wave fields and directions can be found in 
Eq.~(48) of \cite{Barchas-2021-MNRAS}.  It is routinely integrated over solid angles 
\teq{\Omega_f} for the scattered wave to yield the total cross section:
\begin{equation}
   \sigma \; =\; \dover{8\pi}{3} \, r_0^2 \, \dover{ \alphavec \cdot  \alphavec^{\ast} }{ (\omega^2 - \wcyc^2 )^2}  
   \; =\; \dover{\sigt}{ (\omega^2 - \wcyc^2)^2}
      \biggl[ \omega^4 + \wcyc^2 (\wcyc^2 - 2\omega^2) \Bigl\vert \calEvechat_i \cdot \Bvechat \Bigr\vert^2 
      + \omega^2 \wcyc^2 \Bigl\vert \calEvechat_i \times \Bvechat \Bigr\vert^2
         + 2i\, \omega^3 \wcyc \, \Bvechat \cdot \bigl( \calEvechat_i \times \calEvechat_i^* \bigr)
         \biggr] \quad .
 \label{eq:sig_mag_Thom_form}
\end{equation}
The various vector information contained in \teq{\calEvechat_i} and \teq{\Bvec} describes 
the polarization configuration of the incident wave.  For photons
propagating in the \teq{z}-direction we employ the Stokes parameter representation
\begin{equation}
      \calEvec \; =\; {\cal E}_x\hat{x} + {\cal E}_y\hat{y}
   \;\equiv\; \bigl\vert \calEvec \bigr\vert \bigl( \hat{\cal E}_x\hat{x} + \hat{\cal E}_y\hat{y}  \bigr) 
   \quad \Rightarrow\quad
   \boldsymbol{S} \; \equiv\; 
   \left( \begin{array}{c}
   I \\[2pt]
   Q \\[2pt]
   U \\[2pt]
   V
   \end{array} \right) 
   \; =\; 
   \left( \begin{array}{c}
   \langle  {\cal E}_x  {\cal E}_x^* \rangle + \langle  {\cal E}_y {\cal E}_y^* \rangle \\[2pt]
   \langle  {\cal E}_x  {\cal E}_x^* \rangle - \langle {\cal E}_y {\cal E}_y^* \rangle \\[2pt]
   \langle  {\cal E}_x {\cal E}_y^* \rangle + \langle  {\cal E}_x^* {\cal E}_y \rangle \\[2pt]
   i \, \langle   {\cal E}_x {\cal E}_y^* -   {\cal E}_x^* {\cal E}_y \rangle
   \end{array} \right) \quad .
 \label{eq:Stokes_polar_def}
\end{equation}
Here, \teq{\hat{\cal E}_x = {\cal E}_x / \vert \calEvec \vert }
and \teq{\hat{\cal E}_y = {\cal E}_y / \vert \calEvec \vert }, and
the brackets \teq{\langle \dots \rangle} signify time averages of the products of wave 
field components.   For particular surface locales, in Sec.~\ref{sec:local_atmos}
the coordinate axes are chosen using a reference plane coinciding 
with the local zenith axis \teq{\hat{\boldsymbol{n}}} and the photon propagation direction \teq{\kvechat}.
Specifically, \teq{\hat{y} = \hat{\boldsymbol{n}} \times \kvechat/\vert\hat{\boldsymbol{n}} \times \kvechat\vert}
and \teq{\hat{x} = \hat{y} \times \kvechat}.   While this generates non-zero \teq{U} for 
individual photons, integrations over azimuthal angles around the zenith give \teq{U=0} on average 
due to the symmetry of scatterings relative to the \teq{ \hat{\boldsymbol{n}}}-\teq{\boldsymbol{B}} plane.
For implementations involving extended surface regions, the focus of Section~\ref{sec:extended_atmos}, 
the Cartesian coordinates are instead coupled to the rotation axis \teq{\Omegavec} and the 
observer direction \teq{\kvechat_{\infty}}. In such circumstances, \teq{U} is generally not zero.  

\begin{flushleft}
\bf{Appendix B: Field Geometry, Intensity and Ray Tracing}
\end{flushleft}

In this Appendix, we summarize two core general relativistic elements that impact 
the results presented in this paper -- spacetime distortion 
of the magnetic field morphology and ray tracing for curved photon trajectories.   
Here we employ the general relativistic form of a vacuum dipole magnetic field 
in a Schwarzschild metric is detailed in \cite{WS83,Muslimov86,GH94}.
For observer frame polar coordinates \teq{(r, \, \theta, \, \phi )} prescribed relative to the 
instantaneous dipolar axis \teq{\muvechat} (to be distinguished from the local polar coordinates 
used above for the photon atmospheric transport), the field in the local inertial frame (LIF) 
can be expressed as 
\begin{equation}
   \BGR\; =\; 3\dover{B_p\rns^3}{r_s^3} \, \Psi^3 \biggl\{ \xi_r (\Psi)\,\cos\theta\, \hat{r}
      + \xi_{\theta} (\Psi)\,\sin\theta\, \hat{\theta} \biggr\}
      \quad ,\quad
   \Psi\; =\; \dover{r_s}{r}\; \equiv\; \dover{2GM}{c^2r} \quad ,      
 \label{eq:dipole_GR}
\end{equation}
at radius \teq{r}, where \teq{r_s=2GM/c^2} is the Schwarzschild radius of a neutron star 
of mass \teq{M}, whose radius is \teq{\rns}.   Thus \teq{\Psi} serves as the parameter for 
defining the redshift of the photon frequency in propagating from the surface to infinity.  
Throughout the paper, we set \teq{M = 1.44M_{\odot}} and \teq{\rns = 10^6} cm,
so that \teq{\PsiS=r_s/\rns \approx 0.425} is the value of the trajectory radial parameter 
\teq{\Psi = r_s/r} at the surface.  The scaling \teq{B_p} represents the flat spacetime value 
of the polar magnetic field at the surface.  The field component functions are defined by
\begin{equation}
   \xi_r (x) \; = \; - \frac{1}{x^3} \left[ \log_e(1-x) + x + \frac{x^2}{2} \right] 
   \quad ,\quad
   \xi_{\theta} (x) \; = \; \frac{1}{x^3\sqrt{1-x}} \left[ (1-x)\, \log_e(1-x) + x - \frac{x^2}{2} \right] \quad ,
 \label{eq:xi_r_theta_def}
\end{equation}
with \teq{\xi_r(x)\approx 1/3} and \teq{\xi_{\theta} (x)\approx 1/6} when \teq{x\ll 1}
in flat spacetime, reproducing the familiar dipole form.  The influence of general relativity 
is to enhance the magnetic field strength in the LIF.  These forms were used 
to compute the field lines depicted in Fig.~\ref{fig:slab_global_geometry}. 
By taking the ratio of the \teq{\hat{\theta}} and \teq{\hat{r}} components in 
Eq.~(\ref{eq:dipole_GR}), it is promptly determined that
\begin{equation}
   \thetaB \; =\; \arctan \left\{ \dover{ \xi_{\theta} (\Psi)}{\xi_r (\Psi)} \,  \tan\theta \right\} 
 \label{eq:thetaB_ident}
\end{equation}
is the local zenith angle for the atmospheric magnetic field direction at any 
position on the neutron star surface, i.e., that indicated in 
the slab geometry on the left side of Fig.~\ref{fig:slab_global_geometry}.
It is routinely discerned that \teq{\xi_{\theta} (\Psi) /\xi_r (\Psi)} is an 
increasing function of \teq{\Psi}, so that the Schwarzschild metric tilts the 
surface fields somewhat towards the local horizon.

In the Schwarzschild metric, for light emerging from the surface at a position 
\teq{\rvecS = \rns\rvechatS } relative to the star's center in a direction \teq{\kvechatS} 
(its normalized wavector), the trajectory of the light ray lies in the plane defined 
by these two vectors.  This plane also contains the unit radial vector \teq{\rvechat_{\infty}} 
from the center of the star to an observer at infinity, where \teq{\kvechat \equiv\kvechat_{\infty} = \rvechat_{\infty}} .   
The shape of the trajectory is specified by the impact parameter \teq{b}, which represents 
the lateral distance offset between \teq{\rvec_{\infty}} and the asymptote of the trajectory at infinity. 
The impact parameter is related to the zenith angle \teq{\theta_z} of the photon in the LIF
(satisfying \teq{0 \leq \theta_z \leq \pi/2}) as it leaves the atmospheric surface via
\begin{equation}
    \Psi_b\;=\; \frac{r_s}{b}\;=\; \frac{ \PsiS\sqrt{1-\PsiS }}{\sin{\theta_z}} \quad ;
  \label{eq:Psib_def}
\end{equation}
see Eq.~(17) of \cite{GH94}. 
Clearly, there is a one-to-one correspondence between impact parameter and emission
zenith angle.  A schematic depiction of these vectors and the trajectory geometry 
is given in Fig.~\ref{fig:GR_geometry}.  The photon trajectory is most conveniently expressed 
using a single-branched elliptical integral for the orbital angle \teq{\etaorb} between 
the photon position vector \teq{\rvec} at \teq{\Psi} and the photon direction at infinity
\teq{\rvechat_{\infty} = \kinftyvechat}; the well-known result \citep[e.g.,][]{Pechenick-1983-ApJ} is
\begin{equation}
   \etaorb (\Psi)\;=\; \int^{\Psi}_{0}\frac{d\Psi_r}{\sqrt{\Psi^2_b-\Psi^2_r(1-\Psi_r)}}
   \; \approx\; \arccos{ \left( \dover{ \cos{\theta -\Psi}}{1-\Psi} \right) }  \quad .
  \label{eq:theta_Psi_ident}
\end{equation}
The exact trajectory is given by the integral form, and the approximation on the right of 
Eq.~(\ref{eq:theta_Psi_ident}) is that obtained by \cite{Beloborodov-2002-ApJ} that is 
accurate to \teq{\lesssim 1\%}, sufficient for our neutron star applications.  
Here, \teq{\theta} is the zenith angle of the photon momentum \teq{\kGR} in the local inertial frame at any radius,
i.e. its angle to the radial direction.  Thus, \teq{\cos \theta = \kGR \cdot \hat{r} /\vert \kGR\vert}, 
which can be evaluated using the form for \teq{\kGR} posited in Eq.~(41) of \cite{Story-2014-ApJ}:
\begin{equation}
   \kGR\; =\; \dover{\vert \kvec_{\infty}\vert}{\Psi_b\sqrt{1-\Psi }}
    \left\{ \sqrt{\Psi_b^2 - \Psi^2 (1-\Psi )}\, {\hat r} + \Psi \sqrt{1-\Psi} \, {\hat \theta} \right\}
    \quad \Rightarrow\quad   
   \cos\theta \; =\; \dover{1}{\Psi_b} \sqrt{\Psi_b^2 - \Psi^2 (1-\Psi )} \quad .
 \label{eq:momentum_LIF}
\end{equation}
As the photon propagates to infinity, \teq{\Psi \to 0} and the polar component
approaches zero.  When \teq{r=\rns}, (\teq{\Psi = \PsiS}) and the value of \teq{\etaS
\equiv \etaorb (\PsiS )} is obtained using \teq{\theta \to \theta_z} in
Eq.~(\ref{eq:theta_Psi_ident}).  Accordingly, there is a convenient coupling of the
zenith angle intensity information from the atmospheric transport simulation and the GR
propagation to infinity.

The GR propagation computation in {\sl MAGTHOMSCATT} was tested by comparing the
simulated photon trajectories with the exact solutions in
Eq.~(\ref{eq:theta_Psi_ident}). For \teq{M = 1.44 M_{\odot}}, the directional precision
at high altitudes was typically better than 0.1 degrees when the emission zenith angle
satisfied \teq{\theta_z \leq 60^{\circ}}  . This error grows if \teq{\theta_z} is
increased, and reaches a maximum around 3.5 degrees (2.6\%) at
\teq{\theta_z=90^{\circ}}. This precision is sufficient for the directional binning of
3\teq{^{\circ}} on the sky adopted in Figure.~\ref{fig:skymap_w0-1} and
\ref{fig:skymap_w0-5}. We note that most emergent photons have small \teq{\theta_z} in
the atmospheric intensity profiles. The cosine relation in
Eq.~(\ref{eq:theta_Psi_ident}) is valid for \teq{\PsiS} smaller than 0.5. This
\teq{\PsiS} value corresponds to a mass around \teq{1.7 M_{\odot}}, above which the
argument of the arccosine function exceeds \teq{-1}.  An alternative light bending
empirical approximation developed by \cite{Poutanen-2020-AA} gives more accurate results
that can be used for stars with larger masses, ranging up to \teq{2 M_{\odot}}. Yet,
above this mass, the error of Poutanen's formula grows rapidly for large emission
\teq{\theta_z}. Both this and the \cite{Beloborodov-2002-ApJ} approximation address the
expected mass range for normal neutron stars.

A second test of the GR ray tracing implementation was made by investigating the visible
stellar surface for an observer at infinity.  For any observer, the visible surface is
larger than for the flat spacetime case due to gravitational light bending, and for
fixed radius it increases with greater stellar mass. The range of photon orbit angles
\teq{\etaS} corresponding to the visible surface can be obtained using the integral in
Eq.~(\ref{eq:theta_Psi_ident}) by setting \teq{\theta_z} to 90\teq{^{\circ}},
corresponding to emission parallel to the atmospheric horizon. Using the cosine relation
one can quickly obtain the maximum orbit angle \teq{\eta_{\rm max}} such that \teq{0
\leq \etaS \leq \eta_{\rm max}}.  It was determined that the simulation estimate for
\teq{\eta_{\rm max}} agreed with the evaluation of the integral in
Eq.~(\ref{eq:theta_Psi_ident}) within an error of around 3.5 degrees (2.6\%) for \teq{M
= 1.44 M_{\odot}}. 

\begin{flushleft}
{\bf Appendix C: Numerical integration test for intensity and Stokes parameters}
\end{flushleft}

As an independent check on the GR propagation modules in the simulation, the light
curves and Stokes parameters from a neutron star can be calculated using direct
numerical integrations over the entire surface for specified surface anisotropy and
polarization information. We consider a far away observer with a distance
\teq{r\rightarrow\infty} from the star's center. Following \cite{Beloborodov-2002-ApJ},
the observed flux can be expressed in our notations as
\begin{equation}
   dF \;=\; (1-\PsiS) {I_{\omega}}(\muB) \cos{\theta_z}\,
   \frac{d\cos{\theta_z}}{d\cos\etaS} \frac{dS}{r^2}  \quad.
 \label{eq:flux_integral_diff}
\end{equation}
This equation is equivalent to Eq.~(3.14) of \cite{Pechenick-1983-ApJ}, and can also be
derived from the analysis in Section II of \cite{Riffert-1988-ApJ}.  Here the
\teq{1-\PsiS} factor expresses the frequency redshift. The argument
\teq{\muB=\cos{\thetakB}} of the atmospheric intensity \teq{I_{\omega}} is the cosine of
the angle between the photon momentum and the local magnetic field direction (see
Sec.~\ref{sec:injection}), noting that for this test, the injection is azimuthally
symmetric about the local field {\bf B}.  Also,  \teq{\etaS} is the colatitude of the
emission locale measured relative to the observer direction \teq{\kinftyvechat}, and
\teq{dS=\sin{\etaS}d\etaS d\phik} is an area element on the surface. The \teq{dS/{r^2}}
factor reflects the inverse square flux dilution in flat spacetime. The Jacobian
\teq{d\cos{\theta_z}/d\cos{\etaS}} can be solved numerically by differentiating the
integral in Eq.~(\ref{eq:theta_Psi_ident}), or analytically using Beloborodov's cosine
relation therein, which yields
\begin{equation}
   F \;\approx\; \int (1-\PsiS)^2 {I_{\omega}}(\muB) \cos{\theta_z} \frac{dS}{r^2}
   \quad \hbox{since}\quad
   \dover{d\cos{\theta_z}}{d\cos\etaS} \;\approx\; (1-\PsiS)  \quad .
 \label{eq:flux_integral}
\end{equation}
This comparison flux \teq{F} is integrated over the entire visible area on the surface.
For radiation emitted from the whole stellar surface, the upper limit of the trajectory
orbital angle \teq{\etaS} is \teq{\eta_{\rm max}\approx \arccos{[-\PsiS/(1-\PsiS)]}}.
Thus the integrated flux \teq{F} is just a function of the viewing angle \teq{\theta_v},
which varies for a fixed observer as the star rotates.

Similarly, the normalized Stokes parameters can be expressed in terms of the local 
surface Stokes parameters as
\begin{equation}
   \begin{bmatrix}  Q/I \\ U/I \\ V/I \end{bmatrix} \;{\approx} \; \frac{1}{F}
   \int \begin{bmatrix}  Q \\ U \\ V  \end{bmatrix}d\Omega \;{\approx} \; 
      \frac{1}{F} \int (1-\PsiS)^2 {I_{\omega}}(\muB)     
    \begin{bmatrix} \hat{Q}_{\omega} (\muB) \cos{2\delta}\\ \hat{Q}_{\omega} (\muB) \sin{2\delta} 
    \\ \hat{V}_{\omega} (\muB) \end{bmatrix} \cos{\theta_z}\, \frac{dS}{r^2}  \quad,
 \label{eq:Q_U_V_integral}
\end{equation}
where \teq{\delta} is the angle between the reference frame axis at the observer
position and the emission position reference axis parallel transported to the observer
(see Eq.~\ref{eq:x_y_obs_polarize}). To be precise, \teq{\delta} represents the angle of
rotation from the local \teq{\hat{x}} direction (i.e., \teq{\BGRhat\times\kvechatS}) at
the surface emission point to the \teq{\hat{x}} direction at infinity (i.e.,
\teq{\Omegavec\times\kinftyvechat}); thus it depends on the local magnetic field
direction \teq{\BGRhat} and the viewing orbital angle \teq{\etaS}. Here, the empirical
Stokes expressions for \teq{\hat{Q}_{\omega}} and \teq{\hat{V}_{\omega}} in
Eq.~(\ref{eq:QV_Pi_omega_mu}) are used for the surface Stokes values. Note that
\teq{\hat{U}_{\omega}} is always zero due to our choice of Stokes parameter coordinate
system at the surface, and the mapping of \teq{\hat{Q}_{\omega}} at the surface to a
mixture of \teq{Q} and \teq{U} at infinity is connected to a standard rotation of
coordinates through the angle \teq{\delta}.

The test was executed for the dipole field configuration in the Schwarzschild metric
using Eq.~(\ref{eq:intensity_anisotropy}) and Eq.~(\ref{eq:QV_Pi_omega_mu}) to specify
the intensity and polarization distribution function at each point on the emitting
surface of the star.  Thus, the test first employed the asymptotic solutions in domains
of high opacity inside an atmosphere, with known mathematical character, instead of the
full simulation.  The flux and Stokes parameters accumulated over the whole visible
surface, as measured by a far away observer,  were then numerically integrated according
to Eqs.~(\ref{eq:flux_integral}) and~(\ref{eq:Q_U_V_integral}). The intensity and
polarization signatures obtained from the surface integrations depend on the inclination
angle, viewing angle and the rotational phase of the star.  To complete the test, these
numerical integrations were then compared with a series of simulations with photons
injected using Eq.~(\ref{eq:intensity_anisotropy}) and Eq.~(\ref{eq:QV_Pi_omega_mu}) and
\teq{\taueff = 0} for the entire surface. The injected photons thus undergo no
scatterings in the simulation and are subsequently propagated directly to the observer
through curved spacetime. The flux \teq{F}, and Stokes \teq{Q/I}, \teq{U/I}, and
\teq{V/I} from the simulations were obtained as functions of the instantaneous viewing
angle \teq{\theta_v} for selected photon frequencies.  These functions are in numerical
agreement at the \teq{\lesssim 3}\% level with those obtained from surface integration
in Eq.~(\ref{eq:Q_U_V_integral}), for a wide range of \teq{\theta_v} and \teq{\omega},
thereby validating the general relativistic intensity and polarization propagation
implementation in the simulation code.

\end{document}